\documentclass{aa}
\usepackage{graphicx}

\begin{document}

%\thesaurus{3         % Extragalactic astronomy
%              (09.01.2; % ISM: atom1; % kinematics and dynamics
%               11.19s 
%               11.09.1 NGC 1569; % NGC 1569
%               11.09.2; %1; % kinematics and dynamics
%               11.19 interactions
%               11.09.5; % irregular
%               11.11.1; % kinematics and dynamics
%               11.19.3)} % starburst 

\title{HI distribution and kinematics of NGC~1569}
 
\author{J.M. Stil\inst{1,2}
\and F.P. Israel\inst{1}}

\offprints{J.M. Stil}

\institute{
Leiden Observatory, PO Box 9513, NL 2300 RA Leiden, The Netherlands
\and
Department of Physics and Astronomy, University of Calgary, Calgary, Alberta}
 
%\email{stil@ras.ucalgary.ca}
 
\date{Received 00 March 0000; accepted 00 March 0000}

\abstract{ We present WSRT observations of high sensitivity and
resolution of the neutral hydrogen in the starburst dwarf galaxy
NGC~1569.  Assuming a distance of 2.2 Mpc, we find a total HI mass of
$1.3 \times 10^8\ \rm M_\odot$ to be distributed in the form of a
dense, clumpy ridge surrounded by more extended diffuse HI containing
a few additional discrete features, such as a Western HI Arm and an HI
bridge reaching out to a small counterrotating companion cloud. About
10\% by mass of all HI in NGC~1569 is at unusually high
velocities. Some of this HI may be associated with the mass outflow
evident from H$\alpha$ measurements, but some may also be associated
with NGC~1569's HI companion and intervening HI bridge, in which case,
infall rather than outflow might be the cause of the discrepant
velocities. No indication of a large bubble structure was found
in position-velocity maps of the high-velocity HI. The galaxy as a
whole is in modest overall rotation, but the HI gas lacks any sign of
rotation within $60''$ (0.6 kpc) from the center, i.e.  over most of
the optical galaxy. Here, turbulent motions resulting from the starburst 
appear to dominate over rotation. In the outer disk, the
rotational velocities reach a maximum of $35 \pm 6\ \rm km\ s^{-1}$,
but turbulent motion remains significant. Thus, starburst effects are
still noticeable in the outer HI disk, although they are no longer
dominant beyond 0.6 kpc. Even excluding the most extreme high-velocity
HI clouds, NGC~1569 still has an unusually high mean HI velocity
dispersion of $\sigma_v=21.3\rm\ km\ s^{-1}$, more than double that of
other dwarf galaxies.  \\  }
 
\maketitle

\section{Introduction}

NGC~1569 is a small Im type galaxy at a distance of $\rm\ 2.2 \pm
0.6\ Mpc$ (Israel 1988), and a probable member of the IC 342
group (Huchtmeier et al. 2000). With a maximum optical size of
2.9$'$ (1.85 kpc), NGC~1569 is dominated by the aftermath of a burst 
of star formation. Assuming  a Salpeter IMF with slope 2.35 
and lower and upper mass cut-offs at 0.1 and 120 $\rm M_\odot$
respectively, Greggio et al. (1998) found a star formation rate of 
0.5 $\rm M_\odot\ yr^{-1}$ with little change over the past 
150 Myr. Taking into account the limited size and mass of NGC 1569, 
this is a very high rate (Israel 1980). Although the 
present star formation rate is still high, the most intense starburst 
phase occurred between 5 and 10 Myr ago (Israel \& de Bruyn 1988; 
Vallenari \& Bomans 1996; Greggio et al. 1998). The high star
formation rates imply a type II supernova production corresponding to
a total of $2-3 \times 10^5$ over the past 100 Myr in this small
volume of space.

Outflow of gas from NGC~1569 is inferred from the kinematics of an 
extended system of H$\alpha$ filaments
(De Vaucouleurs et al. 1974, Waller 1991, Heckman et al. 1995, Martin
1998) and from extended X-ray emission (Heckman et al. 1995). NGC~1569
is surrounded by an extended halo of relativistic electrons
emitting synchrotron radiation (Israel \& De Bruyn 1988).
 
NGC~1569 contains three very luminous, compact star clusters
designated NGC~1569 A, B and C.  NGC~1569 A may be a double 
cluster; the mass of its brightest component is about
$3 \times 10^{5}\ \rm M_\odot$ (Ho \& Filippenko 1996, De Marchi 
et al. 1997). Given its mass, luminosity and colour, NGC~1569 A 
will closely resemble a globular cluster in the Milky Way after 
an elapsed time of 15 Gyr.

In a previous paper, we have presented evidence for interaction with 
a massive nearby HI cloud, NGC~1569-HI. This cloud is probably
related to the recent starburst in NGC~1569, whereas the nearby 
dwarf companion UGCA~92 almost certainly is not (Stil \& Israel 1998).

Observations of NGC~1569 in the 21-cm line of atomic hydrogen (HI)
have been presented by Reakes (1980) and by Israel \& Van Driel (1990).  
Their maps show the HI structure to consist mainly of a high 
column-density ridge with three local maxima and an arm-like feature 
extending approximately $3'$ to the west. The maximum rotational velocity 
of NGC~1569 is about $30\ \rm km\ s^{-1}$. Israel \& Van Driel (1990) 
also found a region with a local velocity dispersion comparable to the 
rotational velocity.

\section{Observations and data reduction.}

We observed the HI distribution in NGC 1569 with the WSRT during four 
12-hour runs between November 1989 and January 1990. We sampled a total 
of 160 baselines ranging from 36 m to 2772 m with increments of 18 m. 
The 2.5 MHz passband was sampled by 128 independent frequency channels of
$4.1\ \rm km~s^{-1}$ width, yielding a Hanning-smoothed velocity 
resolution of $8.2\ \rm km\ s^{-1}$. Standard gain and phase corrections 
were applied by the reduction group in Dwingeloo. First inspection of 
the data, flagging of bad data and fourier transformation was done in 
the NEWSTAR package. We produced maps at resolutions of $13.5'' ,27''$ 
and $60''$ by applying gaussian weight functions with respective widths 
in the UV-plane of 1960 m, 843 m and 390 m. Resulting dirty 
maps were imported into the {\small GIPSY} package for further reduction 
and analysis. At each pixel, we subtracted continuum emission by 
fitting a first-order polynomial to a total of 48 line-free channels on 
both sides of the frequency band. We then identified manually in each
channel map those areas containing line emission. Within these areas, 
the {\small CLEAN}ing algorithm (H\"ogbom, 1974)
searched for components and subtracted scaled antenna patterns 
from the map down to levels of 0.5 the r.m.s. noise in a single channel 
map. The r.m.s. noise per channel after Hanning-smoothing is 1.1 mJy
per beam at $27''$ resolution. We allowed both positive and negative 
{\small CLEAN} components in order to minimize a bias for positive 
components at low intensity levels.  As interferometer spacings shortwards 
of 36 m are lacking, structures larger than a few arcminutes 
are increasingly diluted. As the HI emission in individual channel maps
rarely extends that far, this effect can safely be ignored.

\section{Results and analysis}

\subsection{21 cm continuum}
\label{contmap-sec}
% NOTE : continuum map noise : cnv : 0.052857 WU = 0.26 mJy/beam
% NOTE : continuum map noise : _60 : 0.100857 WU = 0.50 mJy/beam

\begin{figure}
\resizebox{\columnwidth}{!}{\includegraphics[angle=-90]{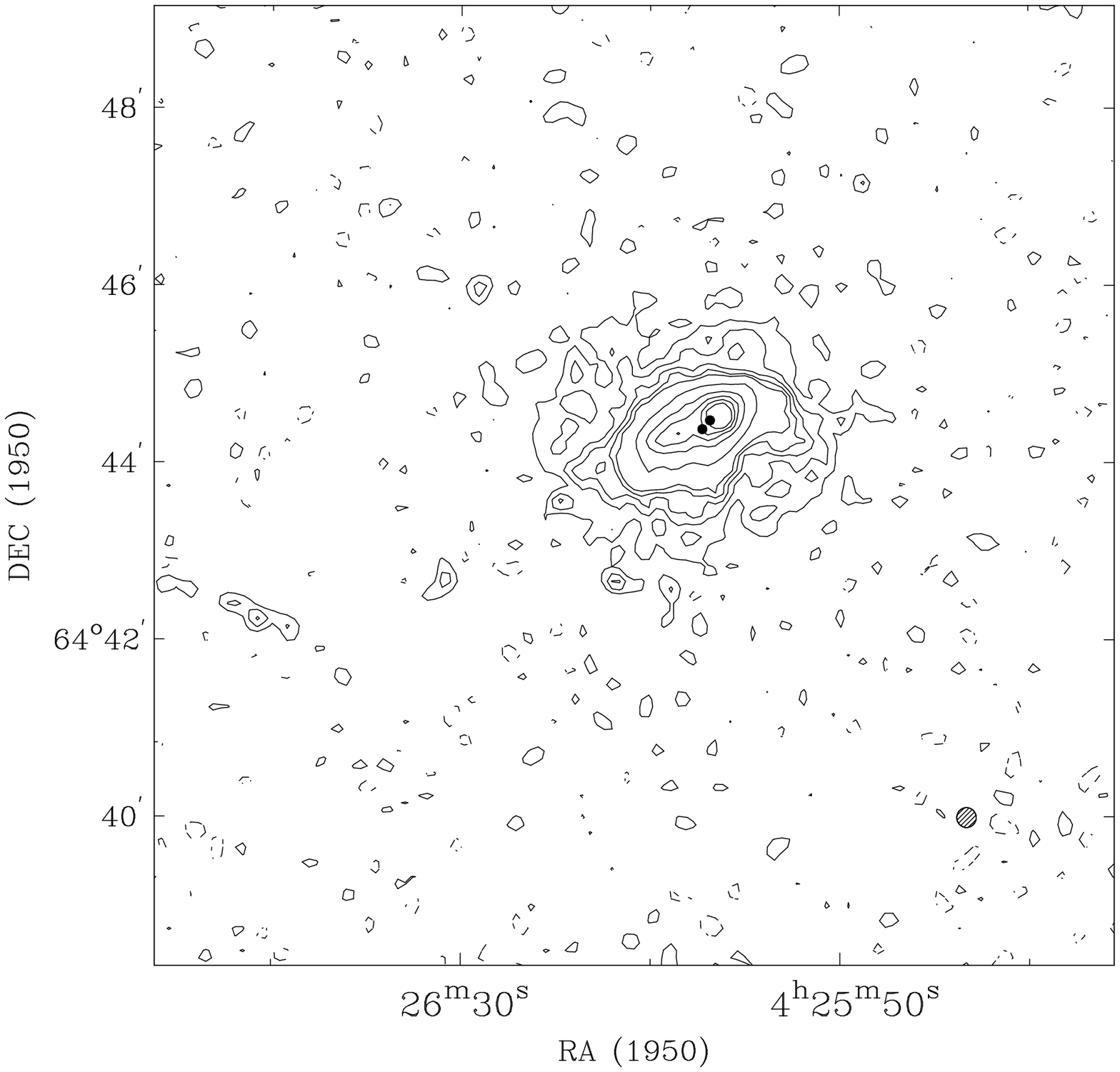}}
\resizebox{\columnwidth}{!}{\includegraphics[angle=-90]{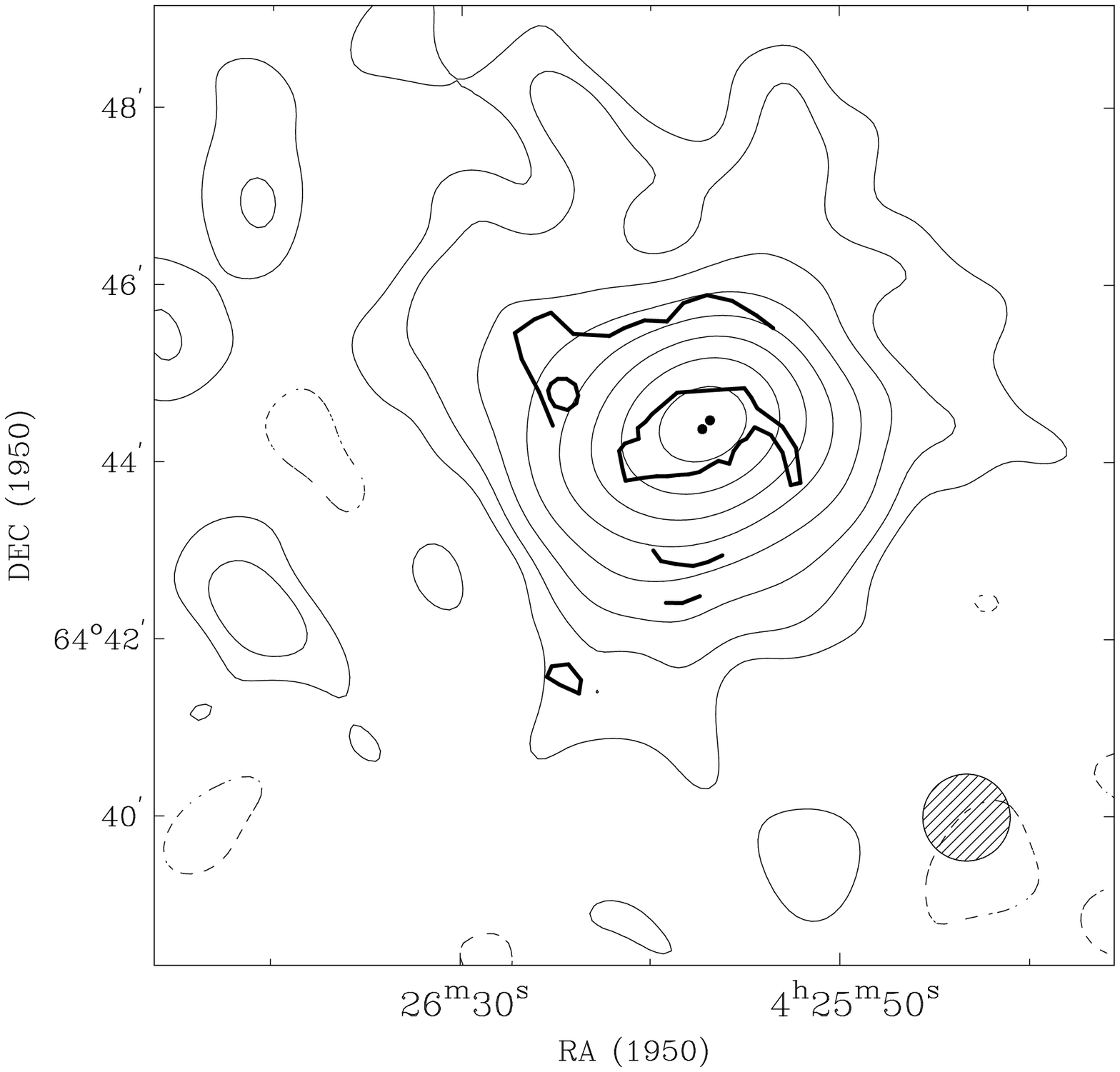}}
\caption{Continuum maps of NGC~1569. Top: resolution $13''$ and contour 
levels at $-0.5$ (dashed), 0.5 (2$\sigma$), 1.0, 1.5, 2.0, 2.5, 3.0, 
3.5, 4.0, 4.5, 5.0, 10.0, 15.0, 20.0, 25.0 mJy/beam. The two dots near 
the center mark the respective locations of super star clusters A and B. 
Bottom: resolution $60''$ and contour levels at -1.0, 1.0 (2$\sigma$), 
2.0, 4.0, 8.0, 16.0, 32.0, 64.0, 128.0 mJy/beam. Clusters A and B are 
marked also in this map as well as the outlines of H$\alpha$ emission.
In both maps, synthesized FWHM beamsizes are indicated by hatched circles.
\label{contmap-fig}
}
\end{figure}

We show the distribution of radio continuum emission in NGC~1569 at 
resolutions of $13.5''$ and $60''$ in Fig.~\ref{contmap-fig}. In the
high-resolution map, the two peaks coincide within 4$''$ with the 
most luminous HII regions labeled 2 and 3 (peak at $\alpha_{1950}=
\rm 4^h 26^m 2^s.7$, $\delta_{1950}=\rm 64^\circ 44' 32''$) and 7 
(peak at $\alpha_{1950}=\rm 4^h 26^m 7^s.0$, $\delta_{1950}=\rm 
64^\circ 44' 20''$) by Waller (1991). The continuum map also shows 
a curved extension at the position of the H$\alpha$ arm. The extended 
radio continuum halo has become visible in the $60''$ resolution map.  
The integrated 21-cm continuum flux-density of NGC~1569 measured in 
the $60''$ resolution map is $438 \pm 5\ \rm mJy$, consistent with 
the results summarized in Israel \& De Bruyn (1988). 

\subsection{21 cm line profile and HI mass}
\begin{figure}
\resizebox{\columnwidth}{!}{\includegraphics{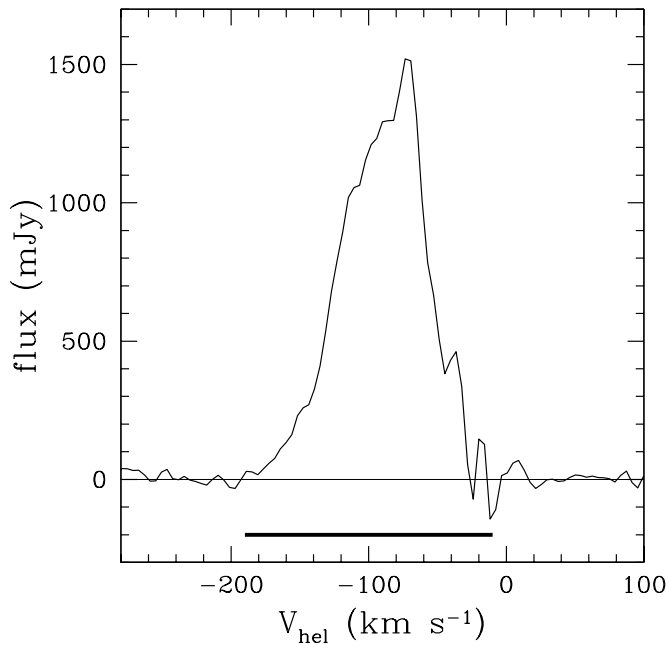}}
\caption{Spatially integrated HI emission line profile of NGC~1569. 
The horizontal line indicates the velocity range over which the line 
flux was integrated.
\label{globalprof-fig}
}
\end{figure}

We obtained the spatially-integrated HI line profile of NGC~1569 
(Fig.~\ref{globalprof-fig})  by summing the flux in each 
channelmap. In order to minimize the effect of sidelobes of the 
un{\small CLEAN}ed Galactic foreground emission, we used
in the summation only those regions that were also selected for 
{\small CLEAN}ing. The asymmetry in the line profile is not due to
absorption by HI in the Galactic foreground, as this would require 
significant optical depths in the velocity range $-70\ \rm km\ 
s^{-1}$ to $-30\ \rm km\ s^{-1}$, which are not observed 
(see Stil \& Israel 1998, their Fig.~4). The HI line flux 
between $v_{\rm hel}=-190~\rm km\ s^{-1}$ and $v_{\rm hel}=-10\ 
\rm km\ s^{-1}$, corrected for primary beam attenuation and 
including NGC~1569-HI (Stil \& Israel 1998) is $116~\rm Jy\ km\ 
s^{-1}$. At the distance of 2.2 Mpc, the HI mass of NGC~1569 is 
thus $M_{\rm HI}=1.3 \times 10^8\ \rm M_\odot$, in good agreement 
with the values listed by Israel (1988).

The intensity-weighted mean velocity of the line profile is $v_{\rm
sys}=-90.3~\rm km\ s^{-1}$, i.e identical to the value
$v_{\rm sys}=-89~\rm km\ s^{-1}$ listed in the RC3 (de Vaucouleurs
et~al. 1991), but approximately $13~\rm km\ s^{-1}$ smaller than the
systemic velocity measured either in H$\alpha$ (Tomita et al. 1994) 
or from the HI velocity field (cf. Reakes 1980). This discrepancy  
probably reflects the asymmetry of the HI line profile (see also
Sect.~\ref{rotcur-sec}).
The presence of small-scale emission from the Galactic foreground 
produces a negative offset of maximum $-0.7$ mJy/beam ($0.5\sigma$) 
in the two most affected channels at $-57\rm\ km\ s^{-1}$ and 
$-16\rm\ km\ s^{-1}$. We have not attempted to reduce the influence 
of Galactic foreground emission by eliminating the shortest baselines,
as this would further dilute any large-scale structures in NGC~1569.

\subsection{Overall HI distribution}
\label{NHImap-sec}

\begin{figure*}
\centerline{\resizebox{\textwidth}{!}{\includegraphics[angle=-90]{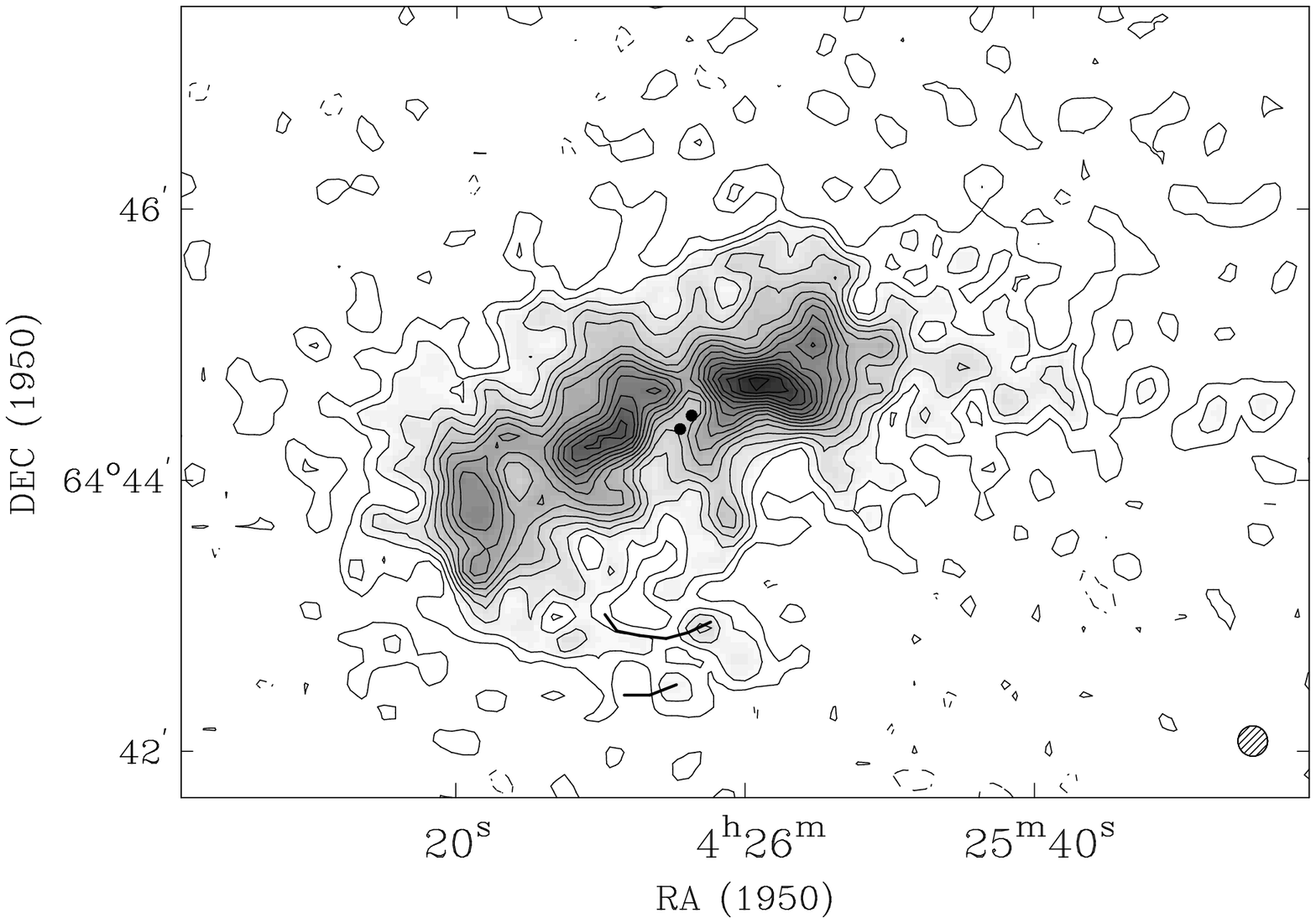}}}
\centerline{\resizebox{\textwidth}{!}{\includegraphics[angle=-90]{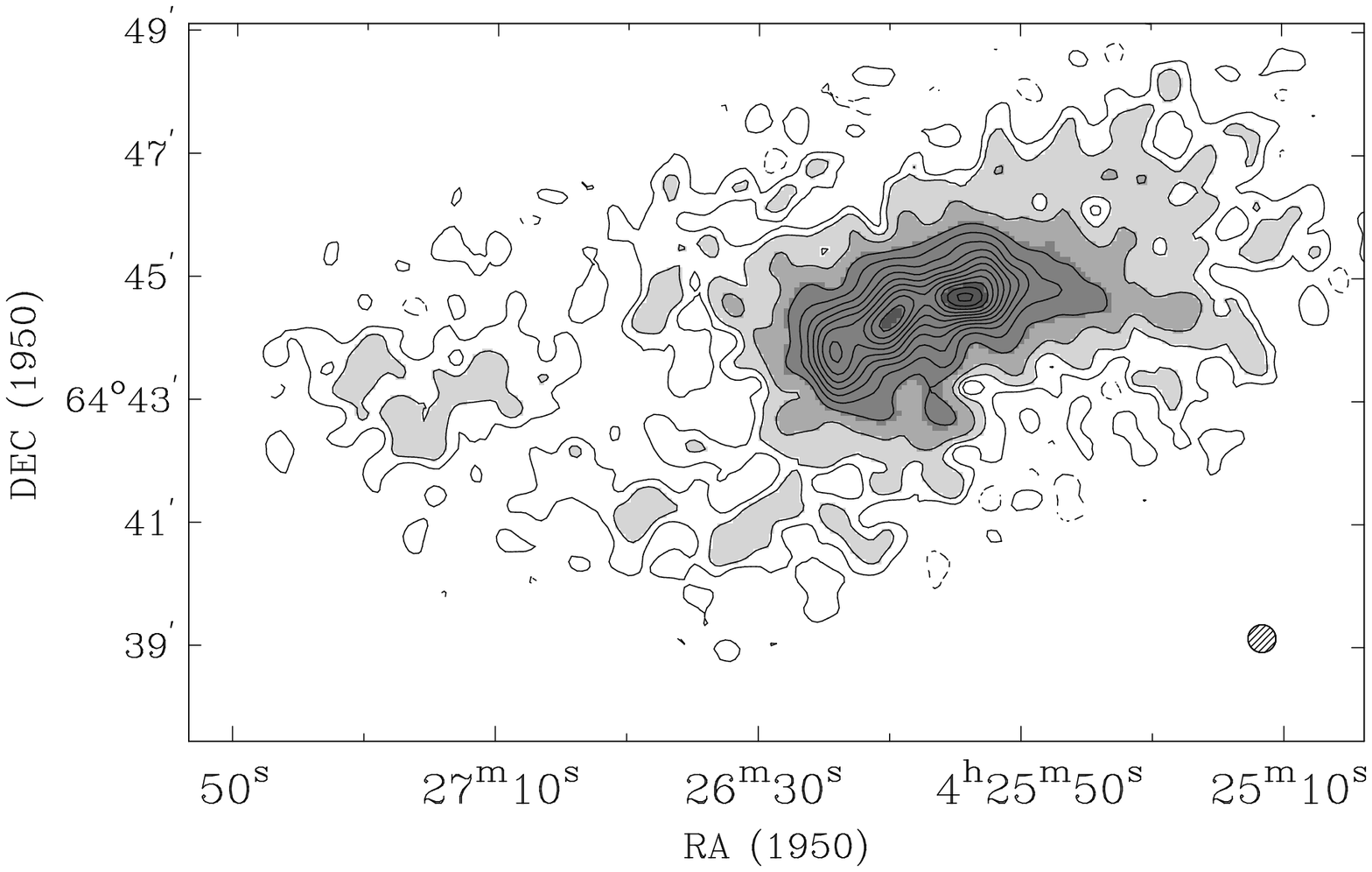}}}
\caption{HI column density maps of NGC~1569. Top: map at $13''$ 
resolution with contours at $-5$ (dashed), $5$, $10$, $15$, $20$, 
$\ldots$ $\times 10^{20}$ HI $\rm cm^{-2}$. Linear grayscales start
at 10.0 $\times 10^{20}$ HI $\rm cm^{-2}$.  The two dots near the 
center mark the position of starclusters A and B. The curved lines 
mark the position of Arc 1 and Arc 2 (Waller 1991). Bottom:
Map at $27''$ resolution with contours at $-1$ (dashed), $1$, $2$, 
$4$, $9$, $14$, $19$, $24$ $\ldots$ $\times 10^{20}$ HI $\rm cm^{-2}$.
Grayscales change at 2.0, 4.0, 8.0, 50.0, and 70.0 $\times 10^{20}$ HI
$\rm cm^{-2}$.  
\label{NHImap-fig}
}
\end{figure*}

Fig.~\ref{NHImap-fig} shows the distribution of HI column 
densities at $13''$ and $27''$ resolution. The low-resolution map 
shows both NGC~1569 and NGC1569-HI (Stil \& Israel 1998). The 
high-resolution map shows only the central region with the 
high column-density ridge. In constructing these maps, 
we applied the areas selected for the {\small CLEAN} algorithm 
as a mask to the data and then integrated the flux over the
observed velocity range. In this way, we both suppress noise 
contributions from empty channels and avoid the bias against 
low column-density components that would be introduced by 
discarding pixels below some intensity threshold.

The overall HI morphology is S-shaped with NGC~1569-HI and the HI
Bridge extending to the southeast. At the other side, a feature which
we will call the Western HI Arm borders a region of low HI
column-densities extending to the north-west. The Western HI Arm
should not be confused with the H$\alpha$ arm, which is much smaller
and nearer to the starclusters A and B. It is not a smooth feature,
but consists of a chain of four to six individual components. The
high-resolution map also reveals the clumpy nature of the main HI
ridge. However, significant amounts of molecular hydrogen probably
fill most of the gap between the two brightest HI maxima (Lisenfeld
et~al. 2002). The optical galaxy is closely associated with this main
ridge. Fig.~\ref{NHImap-fig} shows that the HI hole found by Israel \&
Van Driel (1990) surrounding super starclusters A and B (marked by
dots) connects with a distinct HI depression centered on $\alpha=\rm\
4^h 26^m6^s$, $\delta=64^\circ 44'$. This is bordered in the south by
an HI structure coincident with the H$\alpha$ arcs described by Waller
(1991). Here, the HI bridge apparent in the low-resolution map is
projected onto the main body of NGC~1569. Another, smaller HI minimum
at $\alpha_{1950}= \rm\ 4^h 26^m 15^s.6$, $\delta_{1950}=\rm\ 64^\circ
44' 00''$ ($\pm 5''$ in either direction) appears to be an actual hole
in the HI distribution. At its southwestern edge, a shell-like HII
region centered on a Wolf-Rayet star occurs (Drissen \& Roy 1994).

The large-scale HI morphology and kinematics are reminiscent of tidal
distortion of the outer regions of NGC~1569. However, a major problem
with this interpretation is that this requires a relatively massive,
nearby companion that has not been detected.

\subsection{Velocity field and velocity gradient}

\begin{figure*}
\resizebox{\textwidth}{!}{\includegraphics[angle=-90]{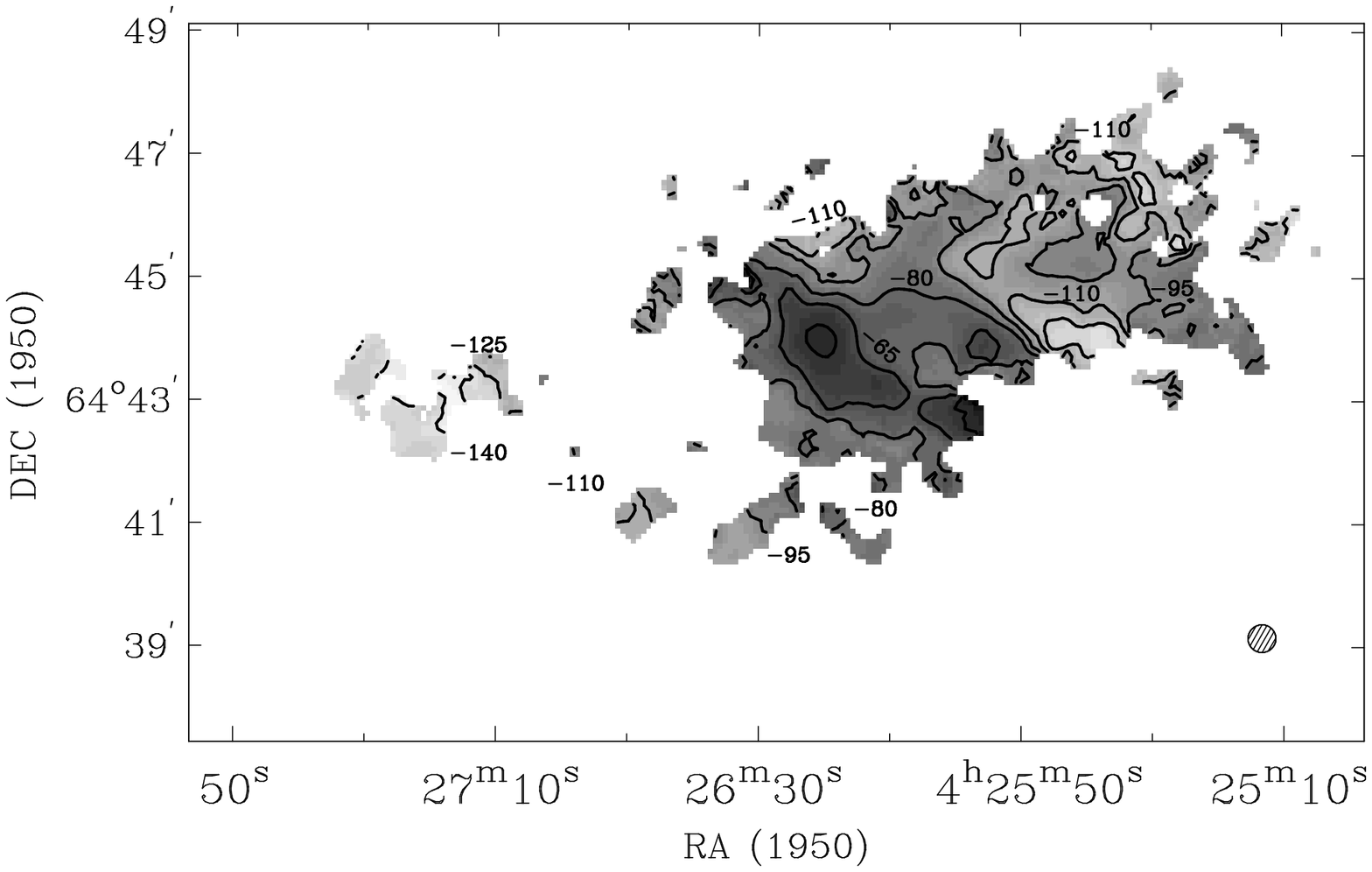}}
\caption{Velocity field of NGC~1569 at $27''$ resolution. Lines of
constant velocity, included at intervals of $15\ \rm km\ s^{-1}$,  
are marked by their heliocentric velocity.
\label{velmap-fig}
}
\end{figure*}

\begin{figure*}
\resizebox{\textwidth}{!}{\includegraphics[angle=-90]{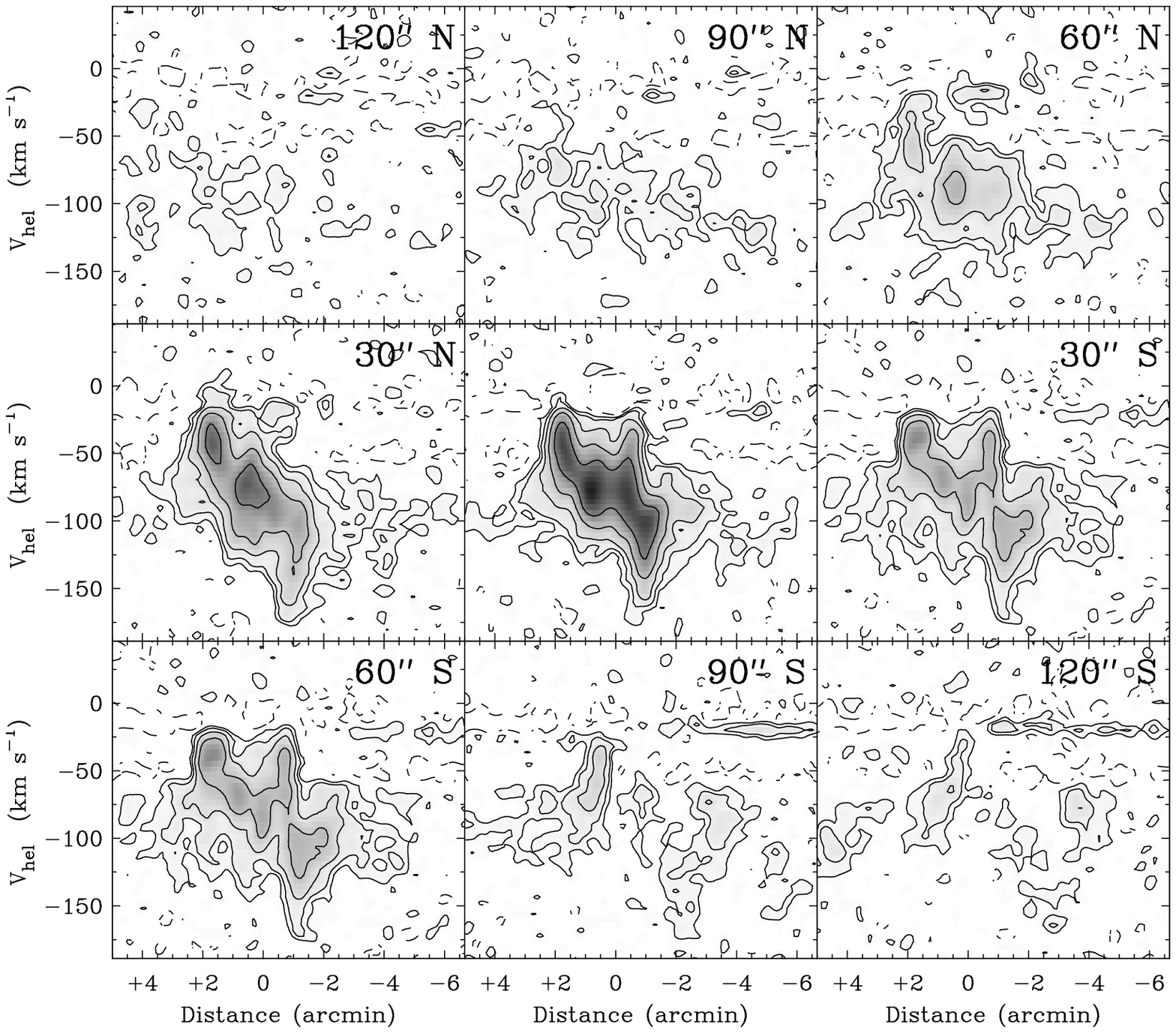}}
\caption{Position-velocity maps along lines parallel to the major axis of 
NGC~1569 (position angle $116^\circ$). The central position of the central 
panel coincides with starcluster A. In all other panels, offsets from the 
major axis are given in the upper right corner. Contours are at $-2.2\ \rm 
mJy$ (dashed), $2.2 \times 2^n\ (n=0, 1, 2, \ldots)$ mJy, linear grayscales 
start at $2.2\ \rm  mJy$.
\label{majXV-fig}
}
\end{figure*}

In Fig.~\ref{velmap-fig}, we show the map of intensity-weighted 
mean velocities. As before, we used the areas selected for the 
{\small CLEAN} algorithm to suppress noise from regions without 
detectable line emission. The overall velocity field clearly shows 
the rotation of NGC~1569 and the apparent retrograde velocity of 
NGC~1569-HI (cf. Stil \& Israel 1998). However, on smaller scales, 
the velocity field exhibits significant local departures from an 
axially symmetric rotating disk. Most of the small-scale structure 
in the velocity field can be traced back to peculiar structures 
visible in individual channel maps. For example, the strongest 
irregularity is a steep velocity gradient near $\alpha_{1950}=\rm 
4^h 25^m 51^s.4$, $\delta_{1950}=\rm 64^\circ 44' 15''$ which
resulted from the superposition of HI in regular rotation and 
an HI cloud with a peculiar velocity readily identified in 
the channel maps and already noted by Israel \& Van Driel (1990)
for its high HI velocity. 
A small east-west velocity gradient is observed in
NGC~1569-HI but it is too chaotic to provide a convincing 
case for rotation. Along the bridge, the velocity changes 
gradually to the velocity of NGC~1569 (Stil \& Israel 1998).

In Fig.~\ref{majXV-fig} we show position-velocity maps 
along lines parallel to the major axis of NGC~1569. Although
starcluster A is not necessarily the dynamic center of 
NGC~1569, it provides a useful reference point and for this
reason was chosen for the zeropoint of the spatial axis in 
the central panel.  The rotation of NGC~1569 is clearly 
visible as the velocity gradient defined by the locus of 
bright emission in the central panel. Note that the region 
of high velocity-dispersion near starcluster A, referred to 
above, is most prominent in the central panel.  We 
will return to this feature below.

The kinematic signature of the HI bridge is apparent in
the p-V diagrams at velocities around 100 km s$^{-1}$ and 
at offsets of $2''$ or more along the major axis. The p-V 
diagrams $90''$ and $120''$ south of starcluster A 
intersect the region of the H$\alpha$ arcs. In these panels,
HI at $1''$ east of starcluster A has a velocity gradient 
opposed to the overall rotation of NGC~1569. This inverted
gradient occurs in a small region where the emission of the
HI Bridge blends into that of NGC~1569. This is also the 
location where the HI bridge might be physically connected 
to NGC~1569, if such a physical connection in fact exists.

\subsection{Rotation curve}
\label{rotcur-sec}

Because of the large departures from circular motion, the 
velocity field of NGC~1569 cannot be analyzed with methods
presupposing regular rotation, such as the method of 
tilted-ring fitting. Before we can accurately determine the 
rotation curve, we must first exclude the high velocity gas 
from the analysis. By fitting ellipses to the regularly shaped
outer isophotes of an R-band image of NGC 1569 taken with the 
1-metre Jacobus Kapteyn Telescope at the Roque de Los Muchachos 
Observatory, we determined the optical center and axial ratio 
of NGC~1569 and used them to define the dynamic center and the 
inclination required for the analysis. The mean center of 
isophotes with semi-major axis between $r=50''$ and $r=75''$
is $\alpha_{1950}=\rm 4^h 26^m 4.^s5$, $\delta_{1950}=64^\circ 
44' 26''$ (r.m.s. scatter of $2''$ in either coordinate). 
The astrometry was tied to the positions of the starclusters 
A and B given by Waller (1991). The position angle of the major 
axis is $117^\circ \pm 2^\circ$ (counterclockwise) with an 
axial ratio $q=0.53 \pm 0.04$. Assuming symmetry around the 
minor axis and an intrinsic thickness $q_0=0.3$, the observed 
axial ratio translates into an inclination $i=63^\circ$.
The shape of the HI distribution, although more poorly defined, 
is consistent with these parameters (see also Reakes 1980),
in particular as the directions of the HI major axis, the 
optical major axis and the overall velocity gradient coincide.

\begin{figure*}
\centerline{\resizebox{\textwidth}{!}{\includegraphics[angle=-90]{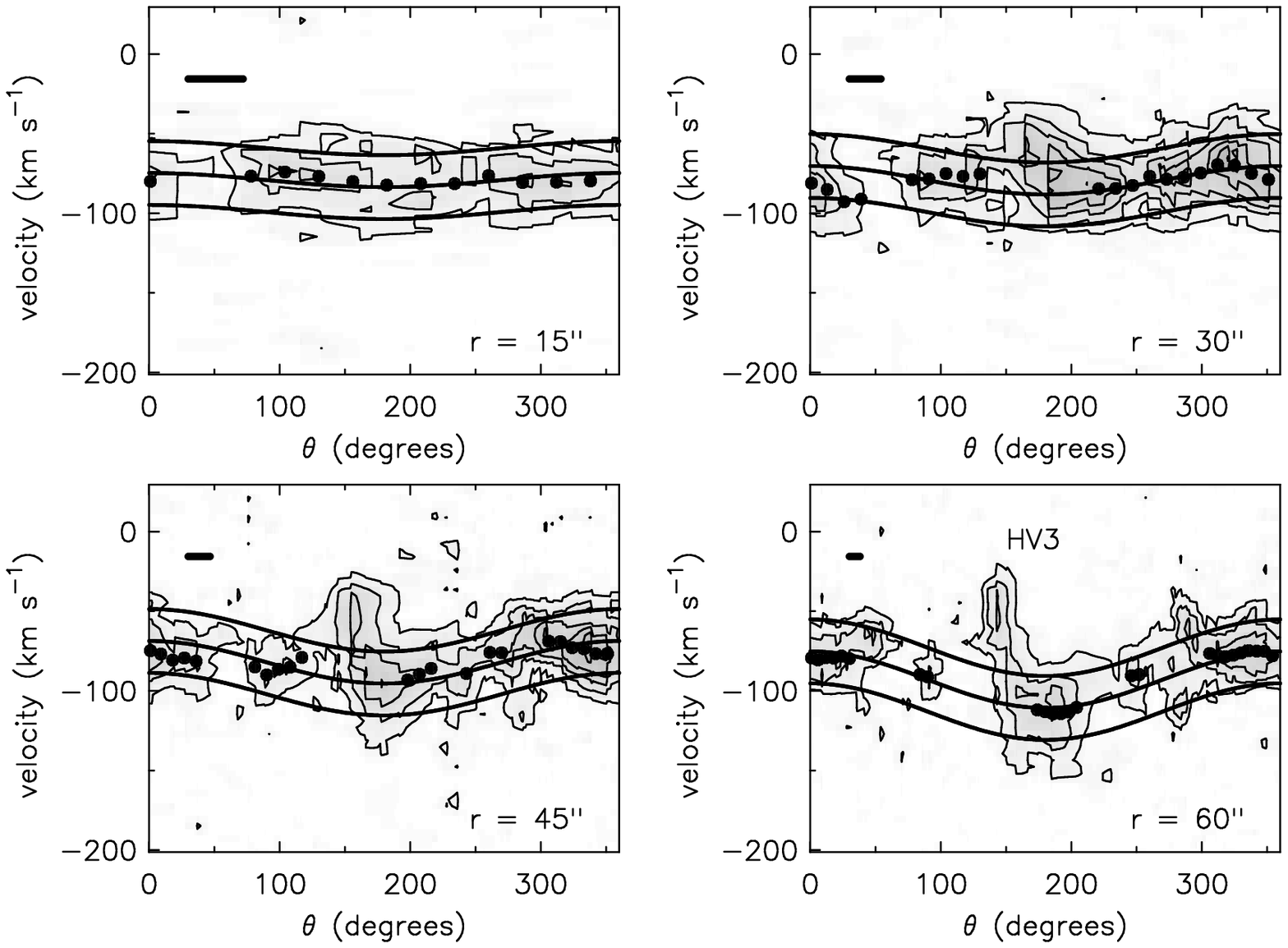}}}
\caption{Position-velocity maps along concentric ellipses defined 
by the optical isophotes. The angle $\theta=0^{\circ}$ corresponds
to the major axis at the receding side (east) of NGC~1569. The 
first four panels ($r \le 60''$) were constructed from the
high-resolution data cube ($13''$ FWHM beam), the remainder from 
the smoothed data ($27''$ FWHM beam). Contours are at positive 
multiples of $5\sigma$, where $\sigma=0.97\ \rm mJy/beam$ (high 
resolution) and $\sigma=1.1\ \rm mJy/beam$ (low resolution) is the 
rms noise in a channel map. Grayscales start at 1.65 mJy/beam. 
A horizontal bar in the first four panels indicates the angle 
occupied by the $13''.5$ (FWHM) beam as seen from the center. 
Dots mark the velocity of gaussians fitted to the line profile. 
The nomenclature of gas with anomalous velocities is defined in
Fig.~\ref{highvprof-fig} and Table~\ref{highv-tab}. Foreground 
Galactic emission is labeled ``G''. The solid curves indicate 
rotation velocities as explained in the text. Figure continued on 
following pages.
\label{ellprof-fig}
\label{ellprof-fig}}
\end{figure*}

\begin{figure*}
\addtocounter{figure}{-1}
\centerline{\resizebox{\textwidth}{!}{\includegraphics[angle=-90]{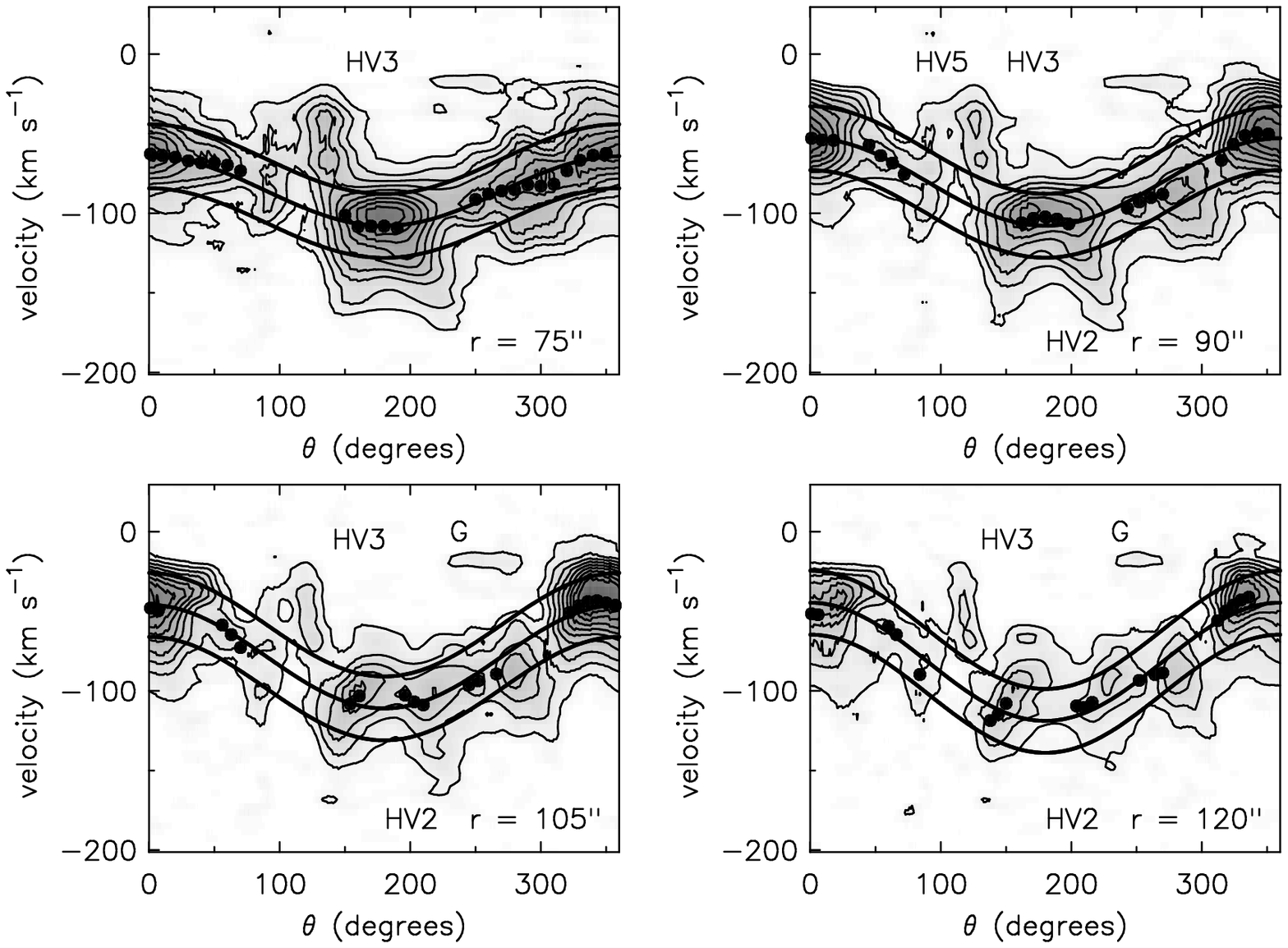}}}
\caption{(continued). 
\label{ellprof-fig}
}
\end{figure*}

\begin{figure*}
\addtocounter{figure}{-1}
\centerline{\resizebox{\textwidth}{!}{\includegraphics[angle=-90]{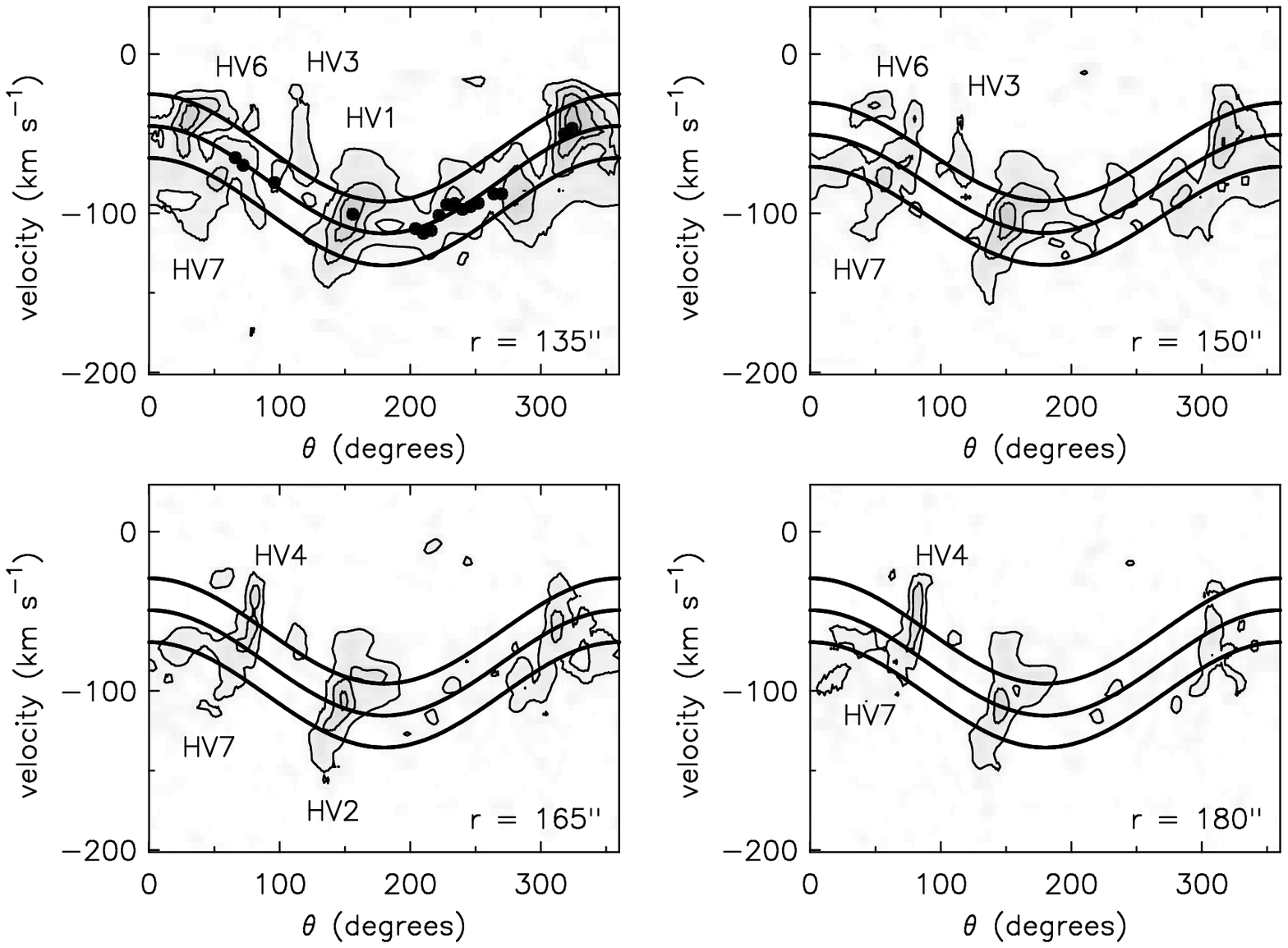}}}
\caption{(continued). 
\label{ellprof-fig}
}
\end{figure*}

\begin{figure}
\resizebox{\columnwidth}{!}{\includegraphics{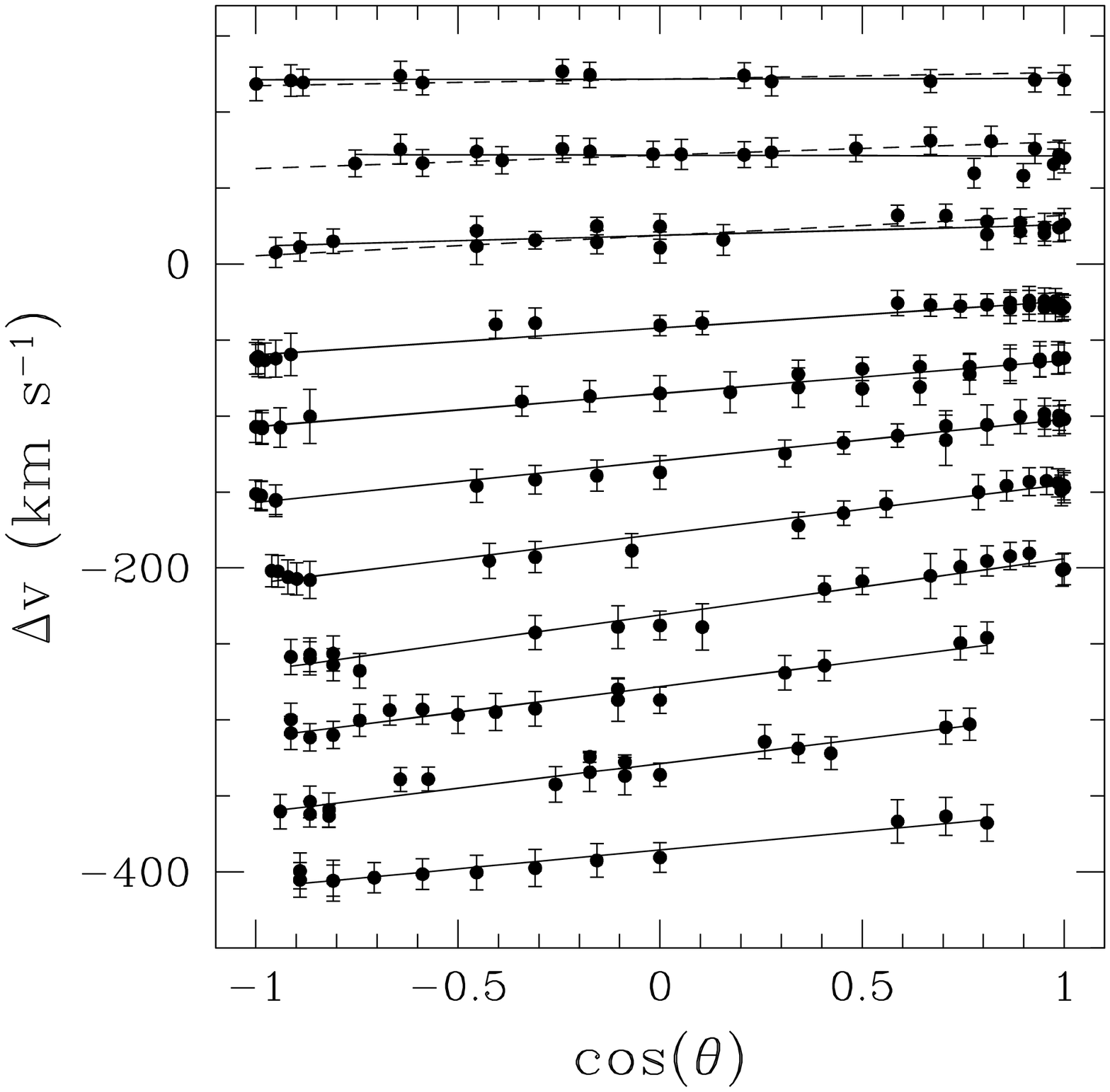}}
\caption{Fits to $v_{\rm obs}=v(r)\sin(i)\cos(\theta)$.  From top to
bottom $r=15''$ to $r=150''$ at $15''$ intervals and $r=180''$ at the
bottom.  For clarity, data for different radii were shifted by $50\
\rm km\ s^{-1}$ relative to the data at $r=75''$. The errorbars
indicate half the fitted velocity dispersion as justified in the
text. The dashed lines indicate the relation expected for solid body
rotation within $r=60''$.
\label{dvcostheta-fig}
}
\end{figure}

To determine the rotation curve of NGC~1569, we followed a procedure 
derived from the classic tilted-ring analysis. We divided the galaxy 
into concentric annuli with the center and axial ratio derived above.
Along these annuli we then constructed p-V maps with the 
{\small GIPSY} task {\small ELLPROF} (Fig.~\ref{ellprof-fig}),
sampling radii ranging from $15''$ to $180''$ at $15''$ intervals.  
For radii $\leq 60''$ we used the high-resolution $13''.5$ in order to
reduce the adverse effects of beamsmearing, while for larger radii we
used the $27''$ data in order to increase the signal-to-noise ratio.

For a single annulus and circular orbits, the observed velocity as a
function of position angle is
$$
v_{\rm obs}(\cos\theta)=v_{\rm sys}+v(r)\sin\,i \cos\theta \eqno(1)
$$
with $v_{\rm sys}$ the systemic velocity, $v(r)$ the rotation velocity
at radius $r$, and the position angle $\theta$ measured in the plane of 
the ring, from the major axis at the receding side of the galaxy. 

The general $\cos\theta$ arrangement of the emission in 
Fig.~\ref{ellprof-fig} confirms the large scale rotation with localized 
but very significant departures from rotation. Peculiar velocities 
similar in magnitude to the rotation velocity itself, are clearly 
visible in Fig.~\ref{ellprof-fig}. A good example is the region of
high velocity dispersion already mentioned, at $r \approx 90''$, 
$\theta=140^{\circ}$. Likewise, the peculiar velocity of the Western 
Arm is clearly visible at radii $r\ge 90''$ between $\theta=170^{\circ}$ 
and $\theta=220^{\circ}$. These and similar features are further discussed 
in Sect.~\ref{highv-sec}.

The velocity $v_{\rm obs}(\cos\theta)$ was measured by fitting a
single gaussian profile to the data at intervals corresponding to
approximately half the FWHM beamsize. For larger annuli, more
independent points are available (see the first four panels in
Fig.~\ref{ellprof-fig}). Areas with clear departures from the
general $\cos(\theta)$ behaviour were avoided.  A least-squares fit of
$v_{\rm obs}$ as a function of $\cos(\theta)$ yielded values of $v_{\rm
sys}$ and $v(r)\sin\,i$ for each annulus. 

\begin{figure}
\resizebox{\columnwidth}{!}{\includegraphics{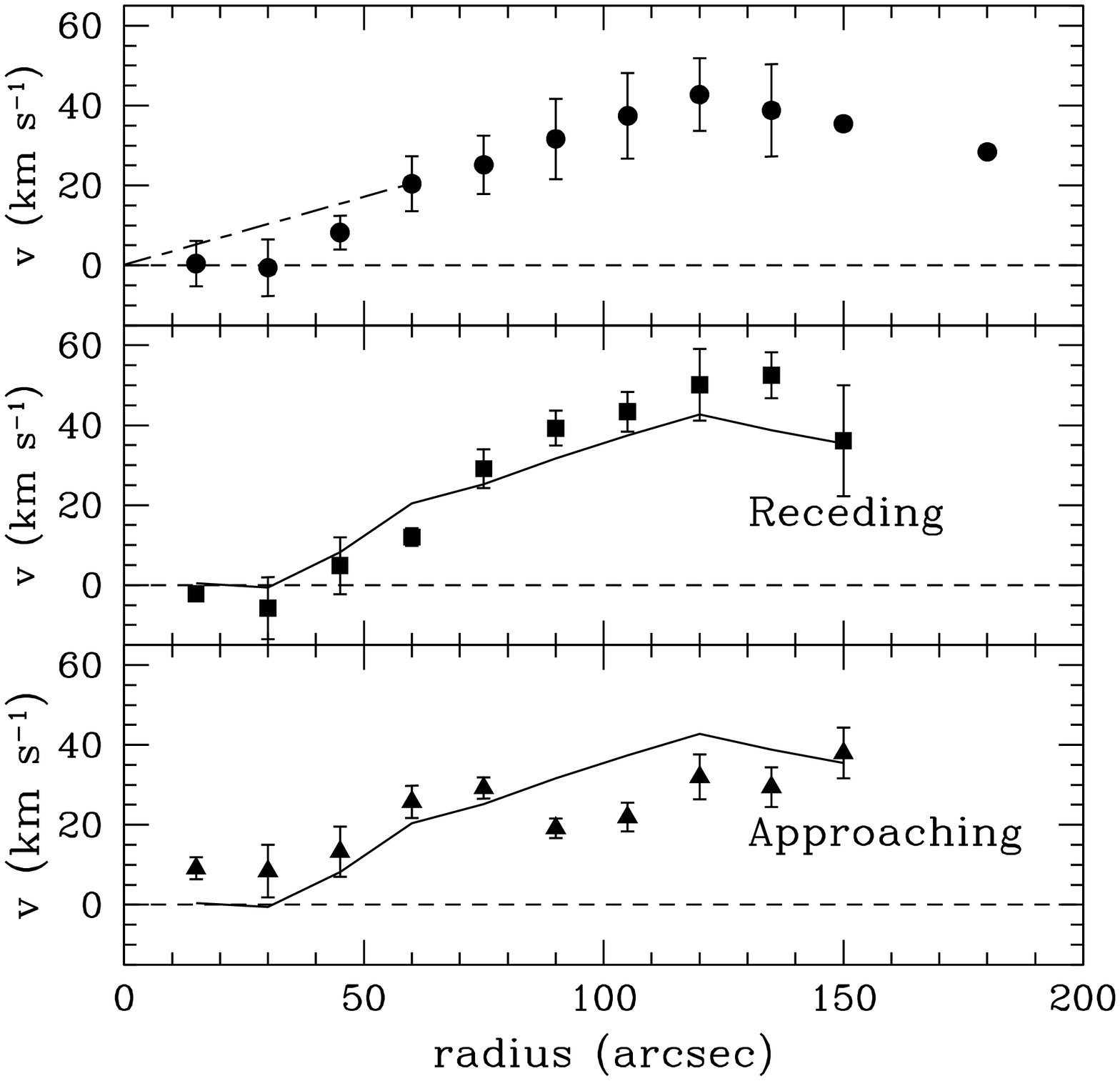}}
\caption{The rotation curve of NGC~1569 fitted to both sides 
simultaneously (upper panel) and to the approaching and
receding sides separately (lower panels). The errorbars in 
the upper panel indicate the difference between the one-sided 
fits (except for $r=75''$, where the mean of the errors at 
$60''$ and $90''$ is used). The dot-dashed line in the upper 
panel indicates solid-body rotation within $60''$ from the 
center. The solid line in the lower panels is the two-sided 
rotation curve from the upper panel. The velocity at radii 
$r \le 60''$ was measured from the $13''$ resolution data, 
and the point at $r=180''$ was measured from the $60''$
resolution data. Intermediate points were measured from 
the $27''$ resolution data.
\label{rotcur-fig}
}
\end{figure}

The fits of equation (1) to the data are shown in Fig.~\ref{dvcostheta-fig}.  
The formal errors in the velocities are a few $\rm km~s^{-1}$. However,
the errorbars in Fig.~\ref{dvcostheta-fig} were defined as half the
fitted velocity dispersion. This is a conservative estimate of the
error in the central velocity introduced by the blending of two
equal-amplitude gaussian components which cannot be separated by
visual inspection.  For radii more than $150''$, the signal-to-noise ratio
and the importance of anomalous velocity components (the Western HI Arm
and the bridge) did not allow an unambiguous determination of the
rotation velocity.  However, a tentative rotation velocity at
$r=180''$ was derived from the $60''$ resolution data.
Table~\ref{rotcur-tab} lists the systemic velocity and the rotation
velocity for the fits shown in Fig.~\ref{dvcostheta-fig}.  The mean
systemic velocity of all radii is $-81.6 \pm 4.4\ \rm km\ s^{-1}$, where
the error is the r.m.s. scatter around the mean.  This is consistent
with $v_{\rm sys}=77 \pm 1\ \rm km\ s^{-1}$ found by Reakes (1980)
from the HI velocity field and by Tomita et~al. (1994) from
H$\alpha$ kinematics, but $9~\rm km\ s^{-1}$ smaller than the 
systemic velocity derived from the 21 cm line profile.  
Separate fits to the receding and approaching side of NGC~1569
are also listed.  The errors in Table~\ref{rotcur-tab} represent
the r.m.s. fit residuals. However, half the velocity difference
between the approaching and receding side is probably a more realistic
estimate of the uncertainty of the rotation velocity. The rotation 
curve thus derived is shown graphically in Fig.~\ref{rotcur-fig}.
The difference between the rotation curves of the approaching and the
receding side is mainly due to limited sampling imposed by the necessity to
avoid contamination by high-velocity HI. 

Rotation velocitiies for $r>60''$ are in good agreement with the
rotation curve of Reakes (1980), which rises linearly with radius to a
maximum of $25\ \rm km\ s^{-1}$ (not corrected for inclination) at $2'$ 
from the center, where it appears to turn over. The subsequent decrease
is, however, only marginally significant. We confirm the suspicion of 
Reakes (1980) that the decline in his rotation curve is due to HI with 
anomalous velocities.

\begin{table*}
\begin{center}
\caption{\bf The rotation curve of NGC~1569 \label{rotcur-tab}}
\begin{tabular}{|  c | r | r |  r |  r  |  l |} 
\hline
r    & $v_{\rm sys}$\ \ \ \ \ \ &$v(r)\sin\,i$&$v_{\rm rec}(r)\sin\,i$ &$v_{\rm app}(r)\sin\,i$ &\ \ \ \ \ \ \ $\sigma_v$ \\
\ [1]   &  [2]\ \ \ \ \ \ &  [3]\ \ \ \ \ \ \ &  [4]\ \ \ \ \ \ \ & [5]\ \ \ \ \ \ \ &\ \ \ \ \ \ [6] \\
arcsec & $\rm km\ s^{-1}$\ \ &$\rm km\ s^{-1}$&$\rm km\ s^{-1}$&$\rm km\ s^{-1}$&\ \ \ \ \ $\rm km\ s^{-1}$ \\ 
\hline
15  & $-78.3 \pm 0.8$   &  $0.4 \pm 1.1$  & $-1.9 \pm 1.3$\  &  $7.9 \pm 2.4$\ & $20.2 \pm 1.0$ \\
30  & $-78.4 \pm 1.5$   &  $-0.5 \pm 2.3$ & $-5.0 \pm 6.7$\  &  $7.3 \pm 5.7$\ & $20.6 \pm 1.4$ \\
45  & $-81.2 \pm 1.1$   &  $7.1 \pm 1.6$  & $4.2 \pm 6.2$\  & $11.5 \pm 5.4$\ & $21.6 \pm 3.1$ \\
60  & $-92.1 \pm 0.8$   &  $17.7 \pm 1.0$ & $10.4 \pm 1.9$\ &  $22.3 \pm 3.5$\ & $21.6 \pm 3.4$ \\
75  & $-85.2 \pm 0.8$   &  $21.8 \pm 1.1$ & $25.2 \pm 4.2$\ &  $25.3 \pm 2.3$\ & $22.6 \pm 4.3$ \\
90  & $-79.5 \pm 1.0$   &  $27.4 \pm 1.3$ & $34.0 \pm 3.8$\ &  $16.6 \pm 2.1$\ & $20.8 \pm 3.4$ \\
105 & $-77.7 \pm 1.0$   &  $32.4 \pm 1.3$ & $37.6 \pm 4.3$\ &  $19.0 \pm 3.1$\ & $21.1 \pm 4.1$ \\
120 & $-80.9 \pm 1.4$   &  $37.0 \pm 2.0$ & $43.4 \pm 7.8$\ &  $27.7 \pm 4.9$\ & $22.0 \pm 3.0$ \\
135 & $-78.1 \pm 1.4$   &  $33.6 \pm 2.2$ & $45.5 \pm 5.0$\ & $25.5 \pm 4.3$\ & $21.2 \pm 3.8$ \\
150 & $-80.7 \pm 1.8$   &  $30.7 \pm 3.3$ & $31.3 \pm 12\ $\  & $32.9 \pm 5.5$\ &\ \ \ \ \ \ \ \ --  \\
180 & $-85.7 \pm 1.2$   &  $24.6 \pm 1.9$ &  --\ \ \ \ \ \ \ \ &  --\ \ \ \ \ \ \ \ &\ \ \ \ \ \ \ \ -- \\ 
\hline 
\end{tabular} 
\end{center}
{\small 
Column definitions:
[1] radius in arcseconds;
[2] systemic velocity in $\rm km\ s^{-1}$;
[3] rotation velocity both sides in $\rm km\ s^{-1}$ (not corrected for inclination);
[4] rotation velocity receding side in $\rm km\ s^{-1}$;
[5] rotation velocity approaching side in $\rm km\ s^{-1}$;
[6] mean velocity dispersion in $\rm km\ s^{-1}$ and r.m.s. scatter around the mean
Notes: The fit of the receding side at $r=45''$ was abandoned because of incoherent
fluctuations of the data; The inconsistency between approaching and receding side
at $r=60''$ is due to the velocities in the interval $300^\circ < \theta < 40^\circ$.
%\label{rotcur-tab}
}
\end{table*}

\begin{figure}
\resizebox{\columnwidth}{!}{\includegraphics{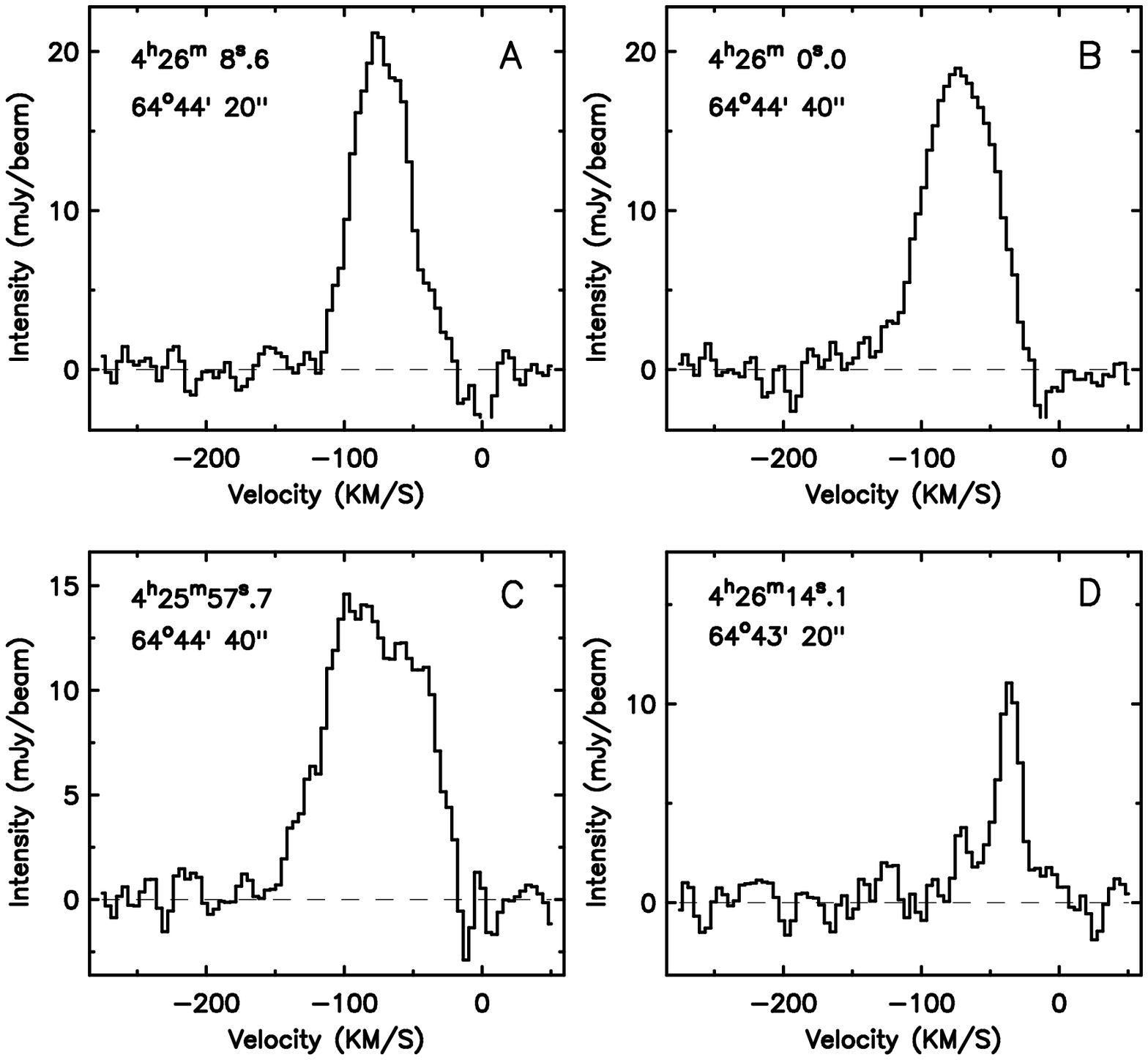}}
\caption{Examples of single-beam line profiles from NGC~1569.  Right
ascension and declination are given in the upper left corner of each
panel. Panels A and B show profiles that can be represented by a
single broad Gaussian. Panel C shows a line of sight with two
components with a velocity difference of $46\ \rm km\ s^{-1}$. Panel D
shows a line profile with a dispersion of $9\ \rm km\ s^{-1}$, which
is observed in most galaxies but is uncommon in NGC~1569.
\label{disp-fig}
}
\end{figure}

If we were to assume a linearly rising rotation curve out to 
radii of $r=75''$, the inferred rotational velocities are 4.5, 9, and 
13.5 $\rm  km\ s^{-1}$ at $r=15''$, $30''$ and $45''$ (see the cosine 
curves in the first four panels of Fig.~\ref{ellprof-fig} and the 
dashed line in Fig.~\ref{dvcostheta-fig}. The actual data at these
radii, however, indicate rotational velocities that are systematically 
smaller. In fact, all rotational velocities for $r \le 60''$ are consistent 
with zero (Figs.~\ref{ellprof-fig} and ~\ref{dvcostheta-fig}). A
lack of rotation in the same region is also evident from 
Fig.~\ref{majXV-fig}.  Beamsmearing is an unlikely cause because such a
discrepancy is not normally found in dwarf galaxies observed at similar 
resolutions (Stil \& Israel 2002). Moreover, a lack of rotation in this
region was also deduced by Castles et al. (1991) and Tomita 
et al. (1994) from H$\alpha$ where beamsmearing plays no role.
Fig.~\ref{ellprof-fig} shows that very low rotational velocities
are representative for all inner HI, and not caused by local high-velocity 
components either. The data are consistent with little or no rotation for 
all values of $\theta$.  

The mean HI velocity dispersion derived from the gaussian fits,
excluding the regions with high-velocity gas, is $21.3\ \rm km\
s^{-1}$.  This is unusually high. In a sample of 26 dwarf galaxies, 
observed at 27$''$ resolution, we find a mean HI velocity dispersion
of $9.5\pm\,0.4 \rm km\ s^{-1}$ (Stil \& Israel 2001). Thus, the
velocity dispersion of the gas in NGC~1569, not even considering
the various clouds with even higher discrepant velocities, already
exceeds that of less disturbed galaxies by more than a factor of two.
Examples of single-beam line profiles, sampling local conditions in
NGC~1569 at $13''.5$ resolution are shown in Fig.~\ref{disp-fig}. 
The contribution of rotation to the width of these profiles is 
negligible and the broad profiles can often be fitted with a single 
gaussian component (Fig.\ref{disp-fig}A, \ref{disp-fig}B). Such 
large velocity dispersions cannot be thermal, as they would then 
imply temperatures $\sim 5 \times 10^4\ \rm K$ characteristic of
ionized rather than neutral hydrogen. Thus, these broad line profiles 
must indicate turbulent velocities greatly exceeding those in  
other galaxies. At some places the line profile is clearly resolved
into multiple components (Fig.~\ref{disp-fig}C) -- note that the 
profiles in \ref{disp-fig}B and \ref{disp-fig}C are separated by only 
$15''$ in right ascension. Locations of more modest velocity 
dispersions $\sim 8\ \rm km\ s^{-1}$ can only occasionally be identified.

\begin{figure}
\resizebox{\columnwidth}{!}{\includegraphics{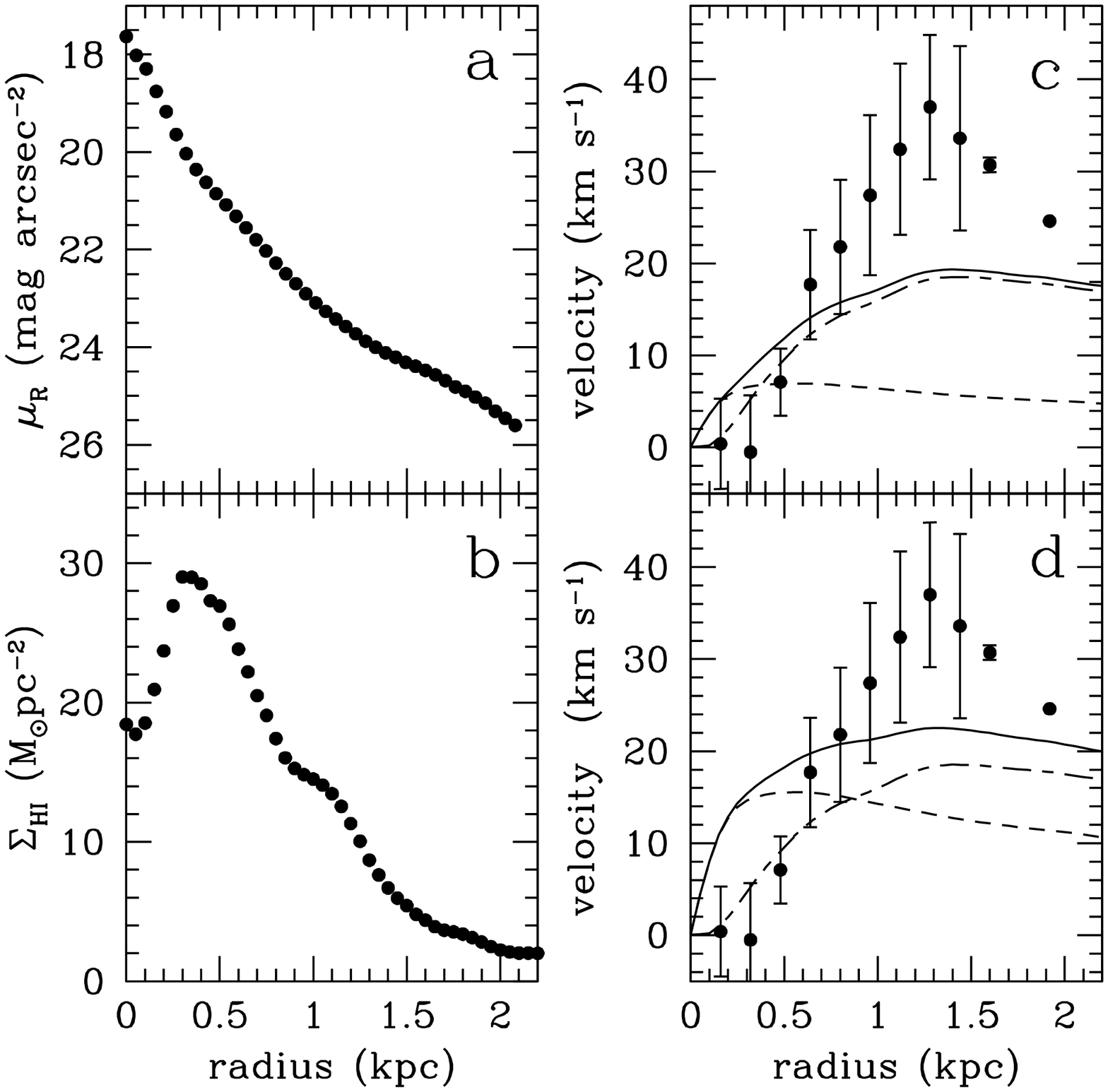}}
\caption{Model rotation curves for NGC~1569. a: R-band surface
brightness profile of NGC~1569.  Foreground stars were
masked out; b: gas surface density, taking into account primordial
helium; c: observed rotation curve compared with model rotation curves
for the stars with $(M/L_{\rm R})_*=0.02\ \rm M_\odot/L_{\rm R \odot}$
(short dashed), gas (long/short dashed) and stars+gas (continuous
line); d: as in c, but with $(M/L_{\rm R})_*=0.1\ \rm M_\odot/L_{R
\odot}$.
\label{massmodel-fig}
}
\end{figure}

No attempt was made to fit a mass model to the rotation curve of NGC
1569, because the asymmetric drift correction introduces a large
uncertainty. However, a model of the rotation curve based on the
visible mass (stars and gas) provides a useful reference frame for the
observed rotation curve and the high velocity HI. In
Fig.~\ref{massmodel-fig} we present model rotation curves for the
observable mass components of NGC~1569 in comparison with the observed
curve. 

We assumed vertical mass distributions for the stars 
and the HI folowing a $\rm sech^2(z/z_0)$ law with scaleheight 
$\rm z_0=0.3\ kpc$. Flatter mass distributions correspond to
steeper rises of the rotation curve.  We normalized the azimuthally 
averaged HI column-density profile to a total mass of $1.7 \times
10^8 \rm M_\odot$ to account for primordial helium. The gaseous mass
does not include $\sim 2 \times 10^6\ \rm M_\odot$ molecular gas
from the bright starformation region west of starcluster A 
(Greve et~al. 1996). We scaled the R-band surface brightness
profile to an integrated extinction-corrected magnitude $R=9.19$ 
($R=10.69$ magnitude from NED and extinction correction in R
based on $E(B-V) = 0.56$ from Israel 1988). We have neglected
any nebular emission line contamination of R-band intensities, 
as the contribution of the H$\alpha$ line to R-band emission
is typically only 10\% to 20\% for an equivalent width of 14.5
nm (Kennicutt \& Kent 1983).  

The model HI rotation curve is comparable to observed rotational
velocities. Inclusion of the stellar rotation curve produces a 
discrepancy between the observed and the predicted rotation curve 
for $(M/L_{\rm R})_* > 0.02$ (Fig.~\ref{massmodel-fig}).
Stellar population synthesis models with $Z=0.004$ and 
steady star formation rates $1\ \rm M_\odot\ yr^{-1}$ predict 
$(M/L_{\rm R})_*$ between 0.05 and 0.1. after $10^8\ \rm yr$, 
depending on the assumed stellar initial mass function 
(Leitherer et al. 1999). The difference between these models 
and the observed rotational velocities imply that turbulent motions
dominate the inner parts.

\subsection{The Western HI Arm}

The Western HI Arm appears as an extension of the high-column-density
ridge in the HI column density map (Fig.~\ref{NHImap-fig}) between
$\alpha=\rm 4^h 25^m 12^s$ and $\alpha=\rm 4^h 25^m 50^s$,
$\delta\approx\rm\ 64^\circ 44'$. It is much more extended than the
system of H$\alpha$ filaments of NGC~1569.  Kinematically, it stands
out because it is redshifted by about $20\rm\ km\ s^{-1}$ from its
surroundings (Figs.~\ref{velmap-fig}, \ref{majXV-fig} and
\ref{ellprof-fig}). On its eastern side, emission of the Arm blends
into emission from the main body of NGC~1569.

It seems unlikely that the Western HI Arm is the symmetric counterpart
of NGC~1569-HI and the HI bridge.  The Western HI Arm is significantly 
closer to the optical center of NGC~1569 than NGC~1569-HI and the HI 
bridge. Furthermore, it has a prograde velocity with respect to the 
rotation of NGC 1569, as opposed to the apparent retrograde 
velocity of NGC~1569-HI (Stil \& Israel 1998).  

\subsection{Neutral high-velocity gas}
\label{highv-sec}
\begin{figure}
%\resizebox{\columnwidth}{!}{\includegraphics[angle=-90]{H357211A.PS}}
%\resizebox{\columnwidth}{!}{\includegraphics[angle=-90]{H357211B.PS}}
\caption{Distribution of the high-velocity HI gas in NGC~1569 in relation 
to the H$\alpha$ emission. Contours are 
at $N_{\rm HI}=-2\ (dashed),\ 2,\ 4,\ 6,\ \ldots\ \times 10^{20}\rm\ 
cm^{-2}$. Components with a negative (approaching) velocity relative
to the model rotation velocity are shown in the top panel. The bottom
panel shows emission with a positive (receding) velocity relative to
the rotation of NGC~1569. Numbers refer to the first column of 
Table~\ref{highv-tab}.  Features 6 and 8 are not considered as real 
and separate high-velocity clouds (see text).
\label{highvmap-fig}
}
\end{figure}

\begin{figure*}
\resizebox{\textwidth}{!}{\includegraphics[angle=0]{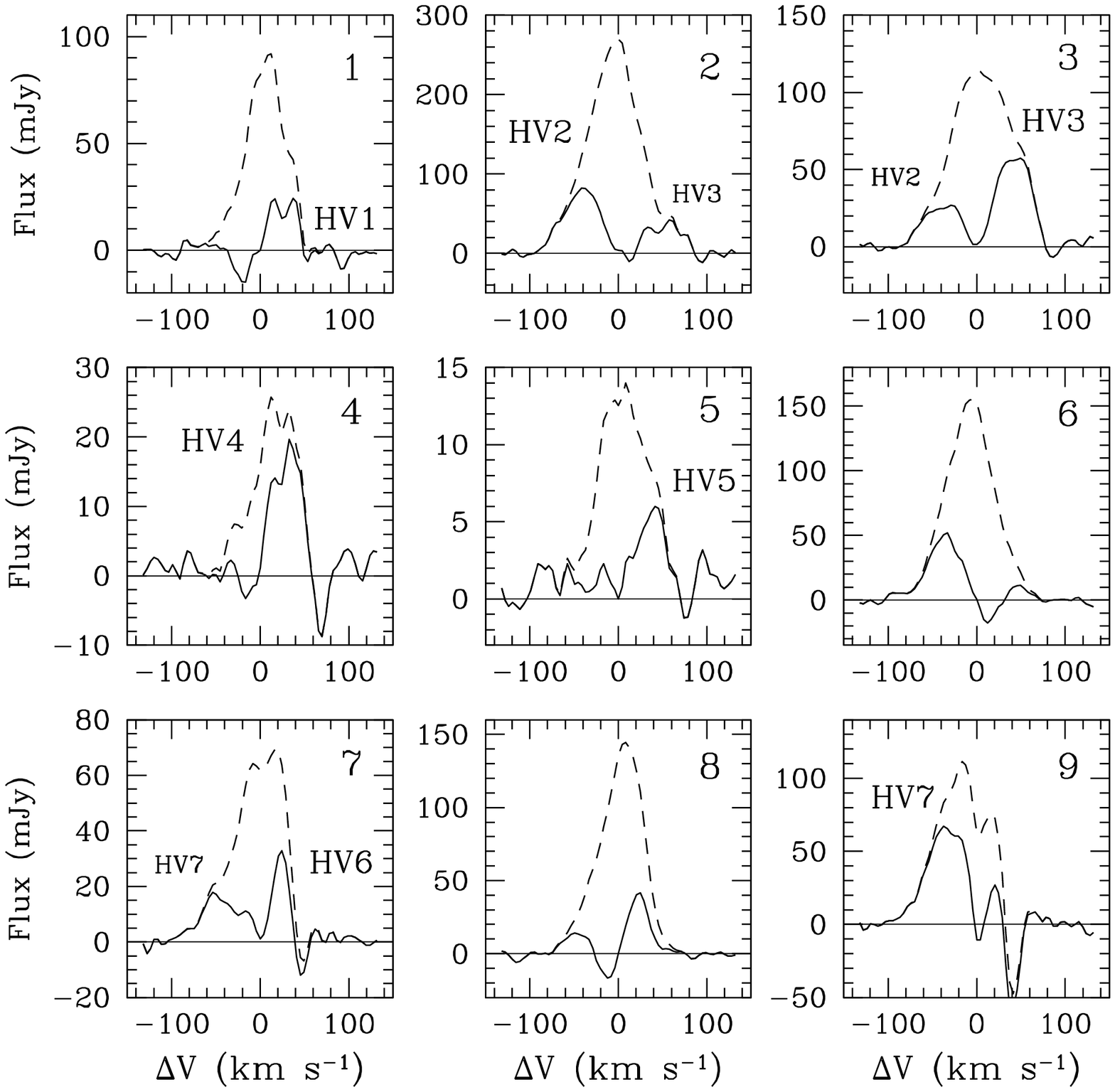}}
\caption{Line profiles of high-velocity components in NGC~1569.  The
dashed line is the line profile before model-subtraction. Solid lines
are the residual line profiles. The zero-point of the velocity axis is
the model rotation velocity.  High velocity components indicated as
HV1 through HV7 can be identified in Fig.~\ref{ellprof-fig}. The panel
numbers refer to column [1] in Table~\ref{highv-tab}.
\label{highvprof-fig}
}
\end{figure*}

Fig.~\ref{ellprof-fig} amply illustrates the presence of local 
but large departures from ordered rotation in NGC~1569.  Attempts
to separate high-velocity gas from the general gas mass by fitting
multiple gaussians to local line profiles do not lead to 
satisfactory results because of insuffiencient signal-to-noise 
ratios outside the HI ridge, complex line profile shapes and
confusion with Galactic HI. Instead, we have isolated the 
high-velocity gas by subtracting a model for the regularly rotating 
gas from the data. To this end, we first constructed a model velocity 
field from the rotation curve in Table~\ref{rotcur-tab}. In 
reasonable agreement with the data in Fig.~\ref{ellprof-fig}, we 
assumed rotational velocities $v(r) \sin\,i=32.5\ \rm km\ s^{-1}$ 
at radii $r \geq 120''$. The model was required to have line-of-sight 
velocities prescribed by the rotation curve and velocity dispersions
equal to the mean dispersion $<\sigma>=21.3\ \rm km\ s^{-1}$. Model
line profile amplitudes are equal to the observed line intensity at 
the velocity predicted by the model rotation curve. We then constructed
HI column-density maps of the residual emission by adding the channel
maps at both positive and negative velocities with respect to the model
rotation velocity. These residual HI column-density maps are shown as 
contours over the H$\alpha$ image in Fig.~\ref{highvmap-fig}. In order 
to suppress noise from regions without significant residual emission, 
only positive and negative residuals stronger than $2 \sigma$ in three 
consecutive velocity channels were included in the summation.  Line 
profiles of the residual components are shown in Fig.~\ref{highvprof-fig}.

The validity of these residuals as high-velocity components was
verified in the channel maps and in Fig.~\ref{ellprof-fig}. The
close correspondence between Fig.~\ref{highvmap-fig} and peculiar
structure in the channel maps adds to our confidence in the
residuals as distinct and separate features (see also 
Fig.~\ref{Hachanmap-fig}). Only the residuals shown in panels 
6 and 8 should be considered as the spurious residuals of bright 
emission with a small peculiar velocity at $r=135''$, 
$\theta=280^\circ$ (residual 6) and $r=120''$, 
$\theta=325^\circ$ (residual 8), and not as separate
components. Component HV1 is associated with the Western HI Arm, 
which is redshifted with respect to its surroundings 
(Fig.~\ref{velmap-fig}).  HV2 is identified as the brightest 
emission in the channel maps at velocities $v_{\rm hel}= 
-168\ \rm km\ s^{-1}$ to $v_{\rm hel}= -135\ \rm km\ s^{-1}$. 
Its  remarkable extension to the southwest is real and clearly 
visible also in the channel maps and in Fig.~\ref{ellprof-fig}
out to radius $r=180''$.  HV2 is just northeast of the H$\alpha$ 
arm and west of the bubble in the north of NGC~1569 designated 
Arc 1 by Tomita et al. 1994) and complex B by Martin (1998).

\begin{table*}
\caption{\bf High velocity components identified in NGC~1569}
\begin{center}
\label{highv-tab}
\small
\begin{tabular}{| c | c |  c |  c |  r |  r |  r | c |} 
\hline 
Panel & Name & $\alpha_{1950}$        & $\delta_{1950}$          & $M_{\rm HI}$          & $v_{\rm hel}$\ \ & $\Delta v$ &log($E_{\rm k}$)  \\
\ [1] & [2]  &  [3]                   &  [4]                     &  [5]\ \ \ \ \         & [6]\ \ & [7]    &   [8]  \\ 
      &      &              &                          & $10^6\ \rm M_\odot$\ \ & $\rm km/s$ & $\rm km/s$  &$\rm J$ \\
\hline 
1     & HV1  &$\rm\ 4^h\ 25^m\ 45^s.5$&$\rm\ 64^\circ\ 45'\ 15''$& $0.4\ \ \pm 0.1$\ \ \ & $-68$  & $37$   &$43.6$  \\
2     & HV2  &$\rm\ 4^h\ 25^m\ 56^s.9$&$\rm\ 64^\circ\ 44'\ 45''$& $4.6\ \ \pm 0.2$\ \ \ & $-139$ & $-43$  &$46.2$  \\
3     & HV3  &$\rm\ 4^h\ 25^m\ 56^s.9$&$\rm\ 64^\circ\ 44'\ 30''$& $3.1\ \ \pm 0.2$\ \ \ & $-51$  & $41$   &$45.3$  \\
4     & HV4  &$\rm\ 4^h\ 25^m\ 59^s.2$&$\rm\ 64^\circ\ 42'\ 45''$& $0.81\pm 0.01$        & $-47$  & $29$   &$44.9$  \\
5     & HV5  &$\rm\ 4^h\ 26^m\ 00^s.0$&$\rm\ 64^\circ\ 43'\ 45''$& $0.26\pm 0.09$        & $-48$  & $35$   &$44.4$  \\
6     &      &$\rm\ 4^h\ 26^m\ 13^s.7$&$\rm\ 64^\circ\ 44'\ 37''$&                       &        &        &        \\ 
7     & HV6  &$\rm\ 4^h\ 26^m\ 15^s.2$&$\rm\ 64^\circ\ 43'\ 15''$&$0.76\pm 0.32$         & $-35$  & $22$   &$45.1$  \\ 
8     &      &$\rm\ 4^h\ 26^m\ 20^s.3$&$\rm\ 64^\circ\ 44'\ 05''$&                       &        &        &        \\  
9     & HV7  &$\rm\ 4^h\ 26^m\ 21^s.8$&$\rm\ 64^\circ\ 42'\ 45''$& $4.0\ \ \pm 0.4$\ \ \ & $-92$  & $-38$  &$44.9$  \\ 
\hline 

Total &      &                        &                          &  $13.9$\ \ \ \ \ \    & $-91$  &        &$46.4$ \\
\hline
\end{tabular}
\end{center}
{\small 
Column definitions :
[1] Identification panel in Fig.~\ref{highvprof-fig};
[2] designation for residuals considered genuine high-velocity components
[3] right ascension (1950) of maximum HI column density;
[4] declination (1950) of maximum HI column density;
%[4] radial range (galactocentric); 
%[5] range in $\theta$;
[5] HI mass assuming a distance of 2.2 Mpc;
[6] heliocentric velocity;
[7] deviation in line of sight velocity from rotation model;
%[7] FWHM velocity width of the  component's line profile
[8] Kinetic energy
}
\end{table*}

Components HV3 and HV4 are visible in the channel maps around $v_{\rm
hel}=-53\ \rm km\ s^{-1}$ and in the $13''.5$ resolution channel map
at $v_{\rm hel}=-48.8\ \rm km\ s^{-1}$ (Fig.~\ref{Hachanmap-fig}).  
The structure of HV3 is remarkably similar to that of the H$\alpha$ 
arm: HI and H$\alpha$ emission have similar extent and display the 
same curvature. The H$\alpha$ emission is only marginally offset to 
the south of HV3 by $\sim 5''$. However, their velocities are very 
different. HV3 is {\it redshifted} relative to the systemic velocity 
of NGC~1569, whereas the H$\alpha$ arm is {\it blueshifted } by 
approximately $40\ \rm km\ s^{-1}$ (Tomita et al. 1994) with respect 
to the systemic velocity. HV5 is separate from HV3, as clearly shown 
in Fig.~\ref{Hachanmap-fig}.
HV6 may be related to HV4; both are found in the direction of
Waller's Arc 1 and together may be its neutral counterpart. The
emission seen in Fig.~\ref{NHImap-fig} (top) at this location is
in fact almost entirely due to HV4 and HV 6.
The location and velocity of HV7 suggest that it is associated with 
the HI bridge between NGC~1569 and NGC~1569-HI.

The parameters of the high-velocity components are listed in
Table~\ref{highv-tab}.  These components account for approximately 
10\% of the HI mass of NGC~1569. Their mean velocity, weighted by 
mass, is $-91\ \rm km\ s^{-1}$, effectively indistinguishable
from the systemic velocity of NGC~1569. This is to be expected 
if the gas was expelled from the galaxy, for instance by the 
starburst. We estimate a lower limit to the kinetic energy of a 
high-velocity component in the restframe of NGC~1569 by 
$E_{\rm k;min} = {1 \over 2} M_{\rm HI} (v-v_{\rm sys})^2$, 
where $v$ and $v_{\rm sys}$ are the observed heliocentric velocity 
of the cloud and of NGC~1569 respectively. These energies are included
in Table~\ref{highv-tab}. The minimal kinetic energy in the 
high-velocity components is $2.5 \times 10^{46}\rm J$. This is 0.3\% 
of the mechanical energy estimated by Heckman et al. (1995) to be
released in the starburst. However, the mass of the neutral 
high-velocity gas is up to an order of magnitude larger 
than the combined mass of the X-ray gas and the H$\alpha$ filaments 
(Martin 1999 and references therein).

\begin{figure}
%\resizebox{\columnwidth}{!}{\includegraphics[angle=-90]{H3572F13.PS}}
\caption{The $13''$ resolution channel map at $v_{\rm hel}=-48.8\ \rm 
km\ s^{-1}$ as contours over the H$\alpha$ emission (grayscale). The 
velocity of the channel map is near the central velocity of 
high-velocity components HV3, HV4 and HV5.  Contours are at 
$-1.94$, $1.94$ ($2\sigma$), $3.88$, $5.82$, $\ldots$ mJy/beam. At this
velocity, HI in regular rotation is expected to be limited to the
area east of the optical galaxy. The spatial coincidence of HV3 and 
the H$\alpha$ arm is clearly visible.  The structure obvious in this 
channel map has a clear counterpart in the velocity field depicted in 
Fig.~\ref{velmap-fig}.
\label{Hachanmap-fig}
}
\end{figure}

\begin{figure}
\resizebox{\columnwidth}{!}{\includegraphics[angle=-90]{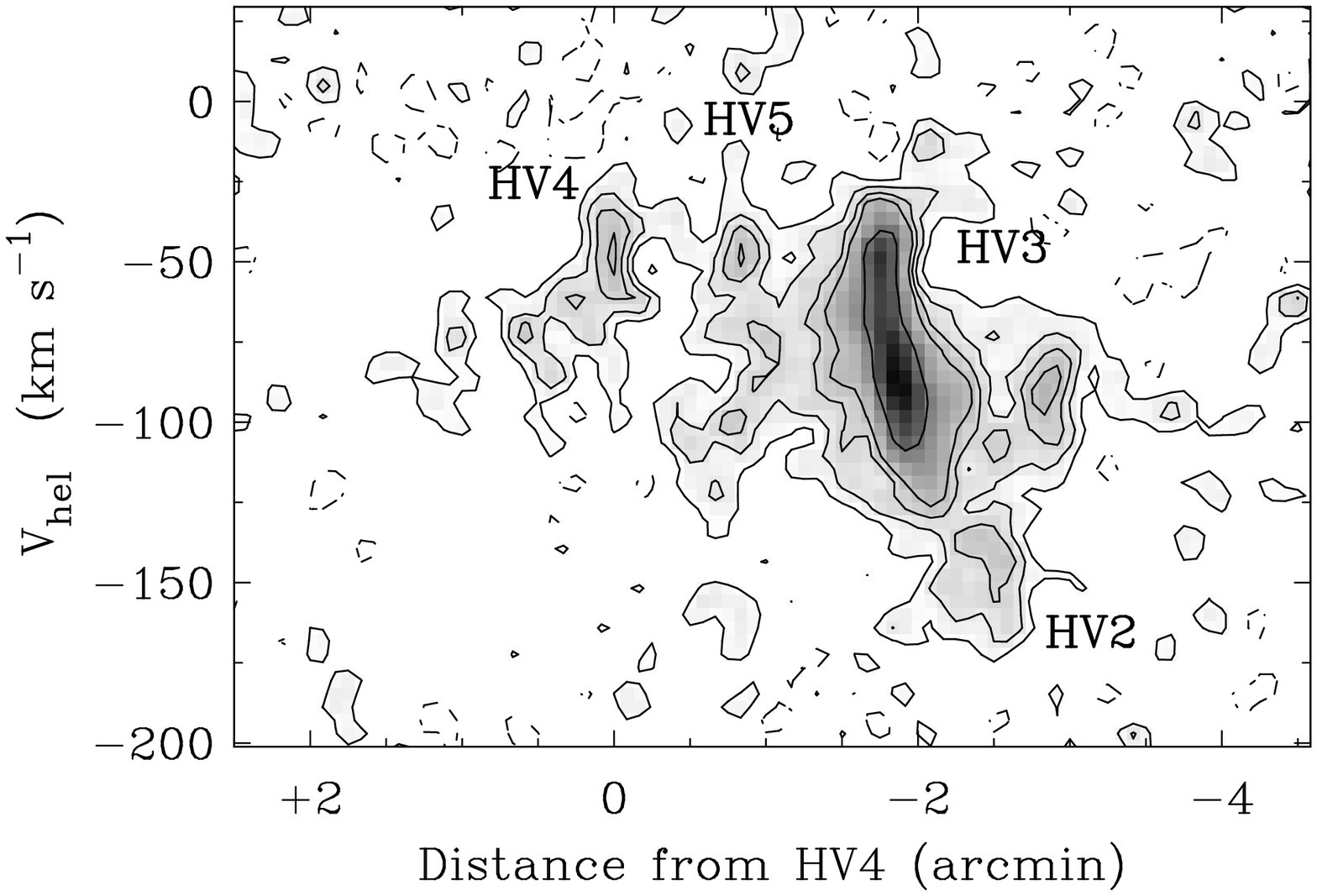}}
\caption{Position-velocity map of the $13''$ resolution data along a 
line through high-velocity components HV3, HV4 and HV5. The only 
candidate doppler-ellipse is the region between HV4 and HV5. This
area coincides with complex G in Martin (1998), which expands with a
velocity of $64\ \rm km\ s^{-1}$. Emission to the left of HV4
(south-southeast) is associated with the HI bridge.  Contours are
at $-1.45$, $1.45$ ($1.5\sigma$), $2.90$, $4.35$, $5.80$,
$11.60$ and $23.20$ mJy per beam.
\label{hv3hv4xv-fig}
}
\end{figure}

The possible association of HV1 and HV4 with H$\alpha$ filaments
raises the question whether the high-velocity components are neutral
gas at the edges of a large bubble as was suggested for the H$\alpha$
filaments (e.g. Waller 1991, Tomita et al. 1994). We looked for evidence
of such a bubble in the form of doppler ellipses combining
several high-velocity components. The position-velocity diagram in
Fig.~\ref{hv3hv4xv-fig} intersects high-velocity components HV2,
HV3, HV4, and HV5. A candidate doppler ellipse is the region between
HV4 and HV5. However, its morphology is quite chaotic.  The faint
emission at the extreme velocities is separated by $70\ \rm km\
s^{-1}$.  If this emission is interpreted as representing the 
approaching and receding edges of an expanding bubble, the implied
expansion velocity of $35\ \rm km\ s^{-1}$ is three times less than 
expansion speeds observed in H$\alpha$ (Tomita et al. 1994, 
Heckman et al. 1995, Martin 1998). The velocity coverage of the present 
HI observations ($-296\ \rm km\ s^{-1}$ to $+116\ \rm km\ s^{-1}$)
includes most of the observed velocities of the H$\alpha$ filaments
(cf. Heckman et al. 1995; Fig.~4). Also, the clumpy ring in 
Fig.~\ref{hv3hv4xv-fig} is even less obvious in p-V maps at 
different position angles. We have thus not found any conclusive 
evidence for a large bubble associated with the high-velocity HI.

Emission from the HI bridge appears to the left (south-southeast) of
HV4 in Fig.~\ref{hv3hv4xv-fig}. There is no trace of the HI bridge
nearer to NGC~1569 than the position of HV4. If the HI bridge does not
terminate at this location, HV4 is the obvious candidate for the
continuation of bridge emission in the line of sight towards NGC~1569.
It is interesting to compare the velocity of HV4 with the velocity of
H$\alpha$ Arc 1.  A long-slit spectrum in position angle $160^\circ$
procured by Heckman et al. 1995 intersects H$\alpha$ Arc 1 somewhat 
to the east of the p-V map in Fig.~\ref{hv3hv4xv-fig}.  If H$\alpha$ 
Arc 1 is the brightest velocity component between $89''$ and $97''$ 
southeast of starcluster A in that spectrum, its velocity of $-75 \pm 
5\ \rm km\ s^{-1}$ is very similar to that of the HI gas southeast of HV4.

\section{Discussion}

\subsection{The rotation curve}

The overall HI kinematics of NGC~1569 are those of a global velocity
gradient along the major axis of the galaxy (Fig.~\ref{velmap-fig}).
This velocity gradient must be caused by rotational movement.  Were it
caused by expansion, we would expect the velocity gradient to be along
the {\it minor} axis, even for radial expansion in the plane of the
disk. The rotation curve of NGC~1569 differs from that of other dwarf
galaxies as its inner velocities are much lower than would be expected
from an extrapolation of the rotation at larger radii. The region of
low velocities, in effect of zero rotation, coincides with the
location of the starburst, and in fact with most of the optically
visible galaxy. Thus, turbulent motions initiated by the starburst
appear to dominate the the inner HI disk remnant. The absence of any
significant decrease in the HI velocity dispersion near $r=60''$
indicates that the ISM is also stirred up in the outer regions of the
galaxy, with a ratio of rotational to random velocities of about
two. Although the starburst did not initiate a global blow-away of the
interstellar medium from NGC~1569, it appears to have left its imprint
on the outer disk. This confirms, at least for NGC~1569, the
conclusions reached by Mac Low \& Ferrara (1999) from hydrodynamic
simulations. Only a partial blow-out seems to occur, even for gas
circular velocities of only $\sim 35\ \rm km\ s^{-1}$.

The rotation curve of NGC~1569 flattens at $90''$ (1 kpc) from the
center, with amplitude $v_{\rm flat}=35 \pm 6\ \rm km\ s^{-1}$ for an
inclination of $60^\circ$.  We attribute the apparent slight decline 
of the rotation curve beyond $120''$ (also observed by Reakes 1980), 
to the peculiar velocities of the HI bridge on the receding side and 
the Western HI Arm on the approaching side. Could the apparent decline 
be caused by warping of the HI disk? The kinematic signatures of 
a warped disk are systematic changes in the position angle of the 
kinematic major axis (dependending on the orientation of the line of 
nodes and the inclination), and changes in the observed rotational 
velocity due to changing inclination. A change in position angle of 
the kinematic major axis would appear as a phase shift of the cosine 
shape in Fig.~\ref{ellprof-fig}. There is no evidence for such a phase 
shift, from which we deduce an upper limit to a change in position 
angle of $20^\circ$. 

\subsection{Neutral high-velocity gas}

Relatively abundant (i.e. 10$\%$ by mass) high-velocity gas has now
been observed in NGC~1569 in both H$\alpha$ and HI- 21cm emission.
The associated HI and HII peculiar velocities are typically 30 to 50
$\rm km\ s^{-1}$ (e.g. Tomita et al. 1994, Heckman et al. 1995, and
Sect.~\ref{highv-sec}).  Doppler ellipses observed in the H$\alpha$
filaments along the minor axis indicate expansion velocities of about
100 $\rm km\ s^{-1}$ (Heckman et al. 1995; Martin 1998), more than
twice the local circular velocity. Such large expansion velocities
that far from the center suggest that at least part of the {\it
ionized} gas associated with the H$\alpha$ filaments may escape from
NGC~1569 (Martin 1998). The temperature of the hot gas is also greatly
exceeds the limit for escape from NGC~1569 (Della Ceca et~al. 1996).
How does the {\it neutral} high-velocity gas relate to the galactic
wind from NGC~1569? This question has important implications for the
ability of the galactic wind to remove chemically enriched material
from NGC~1569 permanently, and the escape of ionizing photons from the
galaxy.

Problems arise if all or most of the high-velocity HI is attributed to
galactic winds blowing out of NGC~1569. The mass of the cold
high-velocity HI components exceeds that of the hot ionized gas by up
to an order of magnitude. This would imply that the gas blown out of
NGC~1569 is largely neutral, contrary to what is expected. The cooling
time of the hot X-ray gas is of the order of $10^9$ years, but this
timescale relates to the hot interior, not the bubble walls (Heckman
et al. 1995). The mass discrepancy is alleviated somewhat if a
fraction of the high-velocity HI should be part of the HI bridge, the
amount of neutral gas to be associated with the galactic wind is
correspondingly smaller. In a previous paper (Stil \& Israel 1998) we
have argued that NGC~1569-HI and the HI bridge are unlikely to be
related to the outflow from NGC~1569. 

The strongest observational evidence linking neutral high-velocity gas 
with the outflow, are the apparent associations of H$\alpha$ Arc 1 
(Waller 1991) with high-velocity components HV4/HV6 and the H$\alpha$ arm 
with HV3. The H$\alpha$ arm is often interpreted as the inner edge 
of a bubble on the southern side of NGC~1569 (Waller 1991, Martin 1998). 
However, in contrast to the spatial coincidence of the H$\alpha$ arm and 
HV3, their radial velocities are very different. This is a significant 
objection because simple radial outflow models cannot simultaneously 
explain blueshifted ionized gas velocities and redshifted neutral gas 
velocities. If both participate in a radial outflow, they should either 
both be blueshifted, or both be redshifted with respect to the systemic 
velocity.  On the other hand, if the H$\alpha$ arm represents the inner
edge of an expanding bubble, most of the expansion velocity will be in 
the plane of the sky, i.e. {\it tangentially}. For sufficiently large 
expansion velocities, a small asymmetry in the ionized and neutral 
gas distributions could lead to a significant difference in the 
(observed) {\it radial} velocities. Nevertheless, the observed radial 
velocity difference of $\sim 40\ \rm km\ s^{-1}$ is almost half the 
maximum expansion velocity inferred from the splitting of the 
H$\alpha$ lines. Yet another possibility is that the dense, relatively 
cool bubble wall is unstable and moves into the hot wind which fills 
the bubble (Suchkov et~al. 1994).

Alternatively, we could consider HV3 and the H$\alpha$ arm to be
part of the same gas system as the HI bridge, and to represent
infall rather than outflow. This gas system would be exposed to the 
galactic winds. It might also be exposed directly to ionizing 
radiation from NGC~1569, as almost half of the ionizing flux produced 
inside the main body of the galaxy escapes (Israel 1988). Heckman et~al. 
(1995) have argued that in the H$\alpha$ filaments photo-ionization 
dominates over shock-heating. The H$\alpha$ arm appears to be 
slightly displaced from HV3 towards the starburst region
(Fig.~\ref{Hachanmap-fig}). However, if the H$\alpha$ arm is
interpreted as the photo-ionized skin of a neutral gas cloud (HV3),
the velocity difference between the neutral and ionized gas remains
unexplained.  Infalling gas colliding with the outflow would cause
shocks that may provide an explanation for the detached X-ray emission
observed by Heckman (1995) and Della Ceca et~al. (1996) in the vicinity
of the H$\alpha$ arcs.

\section{Conclusions}

\noindent
1. The 1.4 GHz continuum flux of NGC~1569 is $438 \pm 5\ \rm mJy$. 
Peaks in the radio continuum distribution coincide with the two
brightest HII regions.  The continuum emission largely follows the 
H$\alpha$ distribution, including the so-called H$\alpha$ arm.

\noindent
2. The HI flux of NGC~1569 measured in the interferometer maps is
$116\ \rm Jy\ km\ s^{-1}$.  This corresponds to an HI mass of $1.3
\times 10^8\ \rm M_\odot$ at 2.2 Mpc. The HI structure of the galaxy 
is that of extended diffuse HI emission, centered on a clumpy ridge
of dense gas associated with the small optical galaxy. Much further
structure is evident, in the form of a counterrotating companion,
and intervening bridge and other features of possible tidal origin.

\noindent
3. There is no detectable rotation of the HI gas within radii of 
$60''$ (0.6 kpc) from the center, confirming earlier results obtained 
for the ionized gas by other authors. Beyond 0.6 kpc from the center, the
rotation velocity increases to $35 \pm 6\ \rm km\ s^{-1}$. An apparent
turnover in the rotation curve is an artifact introduced by the
presence of significant HI with peculiar velocities. Even excluding 
areas with high-velocity HI, NGC~1569 has a high average HI velocity 
dispersion of $\sigma_v=21.3\rm\ km\ s^{-1}$.

\noindent
4.  The starburst has deposited significant amounts of energy into the
disk, creating high velocity features and turbulence such that the
line-of-sight velocity dispersion is much larger than the underlying
rotational velocity due to the mass distribution of the galaxy.  The
starburst has had a more limited effect on the outer disk where
rotational velocities dominate otherwise still significant turbulent
motions.

\noindent
5. In addition, NGC~1569 contains a significant amount of HI gas at
discrepant, high velocities. Components that can be separated from the
regularly rotating gas already provide 10\% of the total HI mass of
the galaxy. At least some of the high-velocity HI appears to be
associated with the H$\alpha$ filaments, but it is unclear whether it
respresents the neutral component of the outflow.  No evidence was
found for a large bubble associated with the high-velocity HI. It is
equally likely that at least part of the high-velocity HI clouds is
associated with the HI companion and bridge linking this companion
with NGC~1569. In that case, infall rather than outflow might be the
cause of the discrepant velocities.

\end{document}